\documentclass[aps,prd,twocolumn,amsmath,amssymb,nofootinbib,longbibliography]{revtex4-2}
\usepackage{amsfonts,amsthm,mathrsfs}
\usepackage{graphicx}
\usepackage[usenames,dvipsnames,svgnames]{xcolor}
\usepackage{times}
\usepackage{inputenc}
\usepackage[normalem]{ulem}
\usepackage{url}
\usepackage{natbib}
\usepackage{subfigure}
\usepackage{booktabs}
\usepackage{cancel}
\usepackage[colorlinks=true,citecolor=blue,urlcolor=blue,linkcolor=blue]{hyperref}

\begin{document}

\preprint{}

\title{Observable Dependence of Viscous Corrections in QGP: Heavy Quarks and Dileptons in Chapman--Enskog Theory}

\author{Lakshmi J. Naik}
\email{jn\_lakshmi@cb.students.amrita.edu}

\author{P. Parvathi}
\email{parvathi20036@gmail.com}

\author{Nachiketa Sarkar}
\email{nachiketa.sarkar@gmail.com}
\author{V. Sreekanth}
\email{v\_sreekanth@cb.amrita.edu}

\affiliation{Department of Physics, Amrita School of Physical Sciences, Coimbatore, \\ Amrita Vishwa Vidyapeetham, India}

\date{\today}

\begin{abstract}
We calculate, for the first time, heavy quark transport and thermal dilepton production from QGP using viscous correction up to second order in gradients. We use the form of viscous correction obtained from Chapman-Enskog like expansion of the Boltzmann transport equation in relaxation time approximation, and compare our results with that of Grad's 14-moment approximation. By employing the temperature and shear stress evolution profiles of QGP obtained from second-order causal relativistic viscous hydrodynamics, we study the heavy quark transport coefficients and thermal dilepton production from an evolving QGP. In the case of HQ transport, the CE corrections suppress the drag force substantially, induce a non-trivial momentum dependence in transverse momentum diffusion, and result in a comparatively less modification in longitudinal momentum diffusion. Whereas, for thermal dileptons, the CE corrections result in an enhanced early-time contribution which decreases and become converging to the first-order CE correction with the evolution of QGP, and remain well behaved compared to that of Grad's correction. Our results indicate that the modification of the observable due to viscous corrections is governed by the magnitude of the corrections as well as the interplay between their momentum dependence and momentum weighting of the transport and emission kernels. 
We demonstrate that 
the momentum structure of the various viscous corrections at the level of distribution function is not directly translated to the observables since the different observables are sensitive to distinct regions of momentum space. 
\end{abstract}

\maketitle

\section{Introduction}
\label{sec:intro}

Ultrarelativistic heavy-ion collision experiments at the Relativitic Heavy Ion Collider, BNL and the Large Hadron Collider, CERN have established the formation of strongly coupled deconfined state of 
quarks and gluons, called the Quark-Gluon Plasma (QGP)~\cite{STAR:2005gfr,PHENIX:2004vcz,BRAHMS:2004adc,PHOBOS:2004zne}. Analysis of the experimental data from these collisions has revealed that the QGP behaves like a fluid with a very low value of shear viscosity to entropy ratio $\eta/s \sim 1/4\pi$~\cite{Kovtun:2004de,Hirano:2005wx,Romatschke:2007mq,Heinz:2013th}. Since the matter produced in these experiments undergoes rapid expansion, only a near local thermodynamic equilibrium can be realized, and non-equilibrium effects can come into existence. These deviations from equilibrium can emerge as several dissipative phenomena such as shear viscous stress, which certainly known to have a significant impact on the evolution of the QGP as well as the observable signatures of the system~\cite{Yagi:2005yb}.  
\par
The evolution of quark-gluon matter created in the heavy-ion collision experiments has been successfully modelled within the framework of causal relativistic viscous hydrodynamics~\cite{Romatschke:2017ejr,Rocha:2023ilf}. The deviations from local thermodynamic equilibrium are encoded through the non-equilibrium corrections to distribution functions. 
The two most commonly used methods in hydrodynamics to determine the form of the near-equilibrium distribution function are the Grad's 14-moment approximation~\cite{Grad:1949zza} and the Chapman-Enskog (CE) method~\cite{Chapman:1970}. In the relativistic variant of Grad's method, the distribution function is expanded as a Taylor series about its local equilibrium value up to second order in particle four-momenta resulting in 14 unknowns~\cite{Israel:1979wp}, and the distribution function is then determined from these coefficients. The Grad's ansatz has been used to study several observables~\cite{Teaney:2003kp,Romatschke:2007mq,Dusling:2007gi,Bhatt:2010cy,Bhatt:2011kx,Das:2012ck}. In contrast, the Chapman-Enskog method constructs the non-equilibrium distribution function as a [gradient expansion] about the local equilibrium distribution~\cite{DeGroot:1980dk}. This method provides a perturbative solution of the Boltzmann equation considering the Knudsen number ($Kn = \lambda/L$, where $\lambda$ is the microscopic mean free path and $L$ is the macroscopic size of the system) as the expansion parameter. 
In Ref.~\cite{Bhalerao:2013pza}, the Chapman-Enskog viscous correction up to second order in gradients was derived by solving the Boltzmann equation iteratively in the relaxation time approximation, thus providing a systematic approach to obtain rapidly converging corrections. The physical consequences of the second-order CE corrections can be examined through observables that are sensitive to the non-equilibrium effects. For this purpose, we consider heavy quarks and thermal dileptons, which are well suited as they directly depend on the phase-space distributions of medium particles.
\par
Heavy quarks (HQs) are one of the efficient probes of QGP, that are mostly produced during the earliest moments of the collision through hard scatterings and carry information about the hotter stages of QGP evolution~\cite{Svetitsky:1987gq,Rapp:2009my,He:2022ywp}. As they have large masses compared to the temperature of the system, once created, they undergo Brownian motion in the hot QGP medium. During their course of propagation through the QGP, they interact with the constituent partons and lose energy, and their dynamics can be described within the Fokker-Planck approach~\cite{Svetitsky:1987gq}. 
The drag and diffusion transport coefficients of HQs contain information regarding their interaction with the medium particles. These transport coefficients are found to be sensitive to various non-equilibrium effects in the medium~\cite{GolamMustafa:1997id,Moore:2004tg,vanHees:2005wb,Berrehrah:2014tva,Das:2015aga,Srivastava:2016igg,Song:2019cqz,Kurian:2020kct,Prakash:2021lwt,Du:2023izb,Chen:2026chk}. 
HQ transport has been analysed in presence of viscous modified distribution functions up to first order in earlier studies involving shear viscosity~\cite{Das:2012ck,Kurian:2020orp,Singh:2019cwi,Shaikh:2021lka,Shaikh:2023qei}. 
\par
Thermal dileptons are another prominent probe of QGP, which are 
emitted throughout the evolution of quark-gluon matter, from different stages~\cite{Salabura:2020tou}. As they interact only electromagnetically, once produced they reach the detectors unaffected. Dileptons offer a distinct advantage owing to their invariant mass $M$, which enables the separation of contribution from different production sources and stages of the collision. Studies on thermal dilepton spectra have provided insights on the effects of different non-equilibrium scenarios in the hot QCD medium~\cite{Bhatt:2011kx,Chandra:2015rdz,Chandra:2016dwy,Kasmaei:2018oag,Coquet:2021lca,Naik:2020jfc,Naik:2022pyk,Coquet:2023wjk,Naik:2025pjt,Garcia-Montero:2024lbl,Wu:2024vyc, Dwibedi:2025xho}. The impact of viscosity has been analysed on thermal dilepton production by considering first-order viscous correction~\cite{Dusling:2008xj,Bhatt:2011kx,Bhalerao:2013aha,Vujanovic:2013jpa,Chandra:2015rdz,Naik:2022pyk,Naik:2025pjt}. 

In the present work, we investigate how the Chapman-Enskog corrections up to second order influence the observables and compare the results with those obtained using Grad's 14-moment approximation, within a common kinetic theory framework. Although various viscous corrections display different hierarchy in momentum at the level of distribution functions, this behaviour is not universal at the observable level. Different observables are sensitive to distinct momentum regions of the medium and consequently receive dominant contributions from different regions of momentum space. This leads to distinct modifications due to the same viscous correction depending on the observable considered. This can be better understood by systematically studying how the corrections propagate at the microscopic level. Understanding how non-equilibrium corrections propagate from microscopic distribution functions to experimentally accessible quantities is of importance in relativistic heavy-ion collisions. A systematic investigation of this observable dependence is therefore essential for establishing reliable connections between dissipative dynamics and measured signals of QGP. 
\par
We consider the heavy quark transport coefficients and thermal dileptons for this purpose as they provide an ideal framework to analyse the phenomenon in a unified way. To the best of our knowledge, this work constitutes the first study to include second-order viscous corrections on heavy quark transport and thermal dilepton production. We look into the momentum structure of the corrections and analyze how they modify the corresponding transport and emission kernels. Effect of viscosity enters to observables not only through underlying distribution functions but also via space-time evolution of QGP as well. 
In this work, the expansion of QGP is modelled using the causal second-order relativistic hydrodynamics within longitudinal boost-invariant Bj\"orken flow~\cite{Bjorken:1982qr}. 
We also examine the spatial diffusion coefficient, which measures the interactions of HQ with medium, in presence of second-order CE corrections. By analyzing both the viscous corrections and their impact on HQ transport and dileptons, we demonstrate that the final manifestation of non-equilibrium effects is governed not only by the magnitude of the viscous correction itself but also by the momentum-space sensitivity of the observable under consideration.

The paper is organized as follows. In Sec.~\ref{sec:deltaf}, we briefly review the Chapman-Enskog and Grad's formalisms and outline the second order causal relativistic dissipative hydrodynamics and Bj\"orken expansion framework employed in the present work. Sections~\ref{sec:heavy-quark} and \ref{sec:dilepton} are devoted to heavy-quark transport and thermal dilepton production, respectively. The numerical results are presented and discussed in Sec.~\ref{sec:results}, followed by a summary and concluding remarks in Sec.~\ref{sec:summary}.

\textit{Notation and conventions}: Throughout the manuscript, we keep $c=\hbar=k_B=1$ and follow the metric convention of $(1,-1,-1,-1)$.



\section{Non-equilibrium corrections and medium evolution}
\label{sec:deltaf}

\subsection{Viscous corrections to distribution functions}   \label{sec:visc-correction}

Presence of viscosity modifies the phase-space distribution functions as 
\begin{equation}
    f_i({\bf p}) = f_i^0({\bf p}) + \delta f_i({\bf p}),
\end{equation}
with $f_i^0$ being the equilibrium part of distribution function of the particle species $i$. The form of viscous correction can be determined from the Boltzmann transport equation. 
The relativistic Boltzmann equation within the relaxation-time approximation is given by~\cite{Anderson:1974nyl}
\begin{eqnarray}
    p^\mu \partial_\mu f_i = -(u \cdot p) \frac{\delta f_i}{\tau_R}; \label{Eq:boltzmann-eq}
\end{eqnarray}
where $\tau_R$ is the relaxation time and $u \cdot p = u^\mu p_\mu$ with $u^\mu$ and $p^\mu$ denoting the four-velocity and four-momenta. We note that $\tau_R$ is independent of momenta and 
$u^\mu$ is defined in the Landau frame~\cite{Anderson:1974nyl}. 
Approximate solutions to the above Boltzmann equation can be obtained using different techniques. 

\subsubsection{Chapman-Enskog like expansion}

In the Chapman-Enskog method, the distribution function is expanded in powers of space-time gradients, about its equilibrium value~\cite{Chapman:1970}, 
\begin{eqnarray}
    \delta f_i = \delta f_i^{(1)} + \delta f_i^{(2)} + \cdots,
\end{eqnarray}
where $\delta f_i^{(1)}$ is first-order in gradients and so on. Rewriting Eq.~\eqref{Eq:boltzmann-eq} in the form $f = f^0 - (\tau_R/(u \cdot p)) p^\mu \partial_\mu f$, the Boltzmann equation can be solved iteratively to obtain
\begin{eqnarray}
    \delta f^{(1)} &=& - \frac{\tau_R}{(u\cdot p)}p^\mu \partial_\mu f^0, \nonumber\\
    \delta f^{(2)} &=& \frac{\tau_R}{(u\cdot p)}p^\mu p^\nu \partial_\mu \left( \frac{\tau_R}{(u \cdot p)} \partial_\nu f^0\right).  \label{Eq:deltaf-iteration}
\end{eqnarray}
Using the above equations in the non-equilibrium part of energy-momentum tensor
\begin{eqnarray}
    T^{\mu\nu} &=& \int \frac{d^3 {\bf p}}{(2\pi)^3 E}  \,p^\mu p^\nu (f + \delta f^{(1)} + \delta f^{(2)})\nonumber \\ 
    &=& \epsilon u^\mu u^\nu - P\Delta^{\mu\nu} + \pi^{\mu\nu}, 
\end{eqnarray}
we obtain the second-order shear stress tensor $\pi^{\mu\nu}$ evolution as~\cite{Jaiswal:2013npa}
\begin{eqnarray}
\dot{\pi}^{\langle \mu\nu \rangle} + \frac{\pi^{\mu\nu}}{\tau_R} &=& 2\beta_\pi \sigma^{\mu\nu} + 2 \pi_\gamma^{\langle\mu} \omega^{\nu\rangle\gamma}\nonumber \\
&& - \frac{10}{7} \pi_\gamma^{\langle\mu} \sigma ^{\nu\rangle\gamma} - \frac{4}{3} \pi^{\mu\nu} \Theta.
\label{Eq:shear-evo-2} 
\end{eqnarray}
In the above expressions, $\epsilon$ and $P$ are the energy density, pressure respectively, $\omega^{\mu\nu} \equiv (\nabla^\mu u^\nu - \nabla^\nu u^\mu)/2$ is the vorticity tensor, $\beta_\pi = (\epsilon + P)/5$, and $\Theta \equiv \partial_\mu u^\mu$. Further, $\sigma^{\mu\nu} \equiv (\nabla^\mu u^\nu + \nabla^\nu u^\mu)/2 - \Delta^{\mu\nu}\Theta/3$, with $\Delta^{\mu\nu} \equiv (g^{\mu\nu} - u^\mu u^\nu)$ being the projection operator orthogonal to $u^\mu$. Also, we use the notation $\dot{A} \equiv u^\mu \partial_\mu A$ and $A^{\langle \mu \nu \rangle}  \equiv \Delta_{\alpha\beta} ^{\mu\nu} A^{\alpha \beta}$, where $\Delta _{\mu\nu}^{\alpha\beta} \equiv \frac{1}{2}(\Delta_\mu^\alpha \Delta_\nu^\beta + \Delta_\nu^\alpha \Delta_\mu^\beta-\frac{2}{3} \Delta^{\alpha\beta} \Delta_{\mu\nu})$.

Now, employing Eq.~\eqref{Eq:shear-evo-2} for $\sigma^{\mu\nu}$ in Eqs.~\eqref{Eq:deltaf-iteration}, we systematically obtain the viscous correction up to second-order in gradients as~\cite{Bhalerao:2013pza}
\begin{eqnarray}
 \delta f^{CE}_i &=& \delta f^{(1)}_i + \delta f^{(2)}_i;
 \end{eqnarray}
 where, 
 \begin{eqnarray}
 \delta f^{(1)}_i&=& \frac{f_i^0 \tilde{f}_i^0 \beta}{2\beta_\pi (u\cdot p)} p^\alpha p^\beta \pi_{\alpha\beta}, \\
 \delta f^{(2)}_i &=&
  - \frac{f_i^0 \tilde{f}_i^0 \beta}{\beta_\pi}\Bigg[
  \frac{\tau_\pi}{(u\cdot p)} p^\alpha p^\beta \pi_\alpha^\gamma \omega_{\beta\gamma}       \nonumber \\
&&-\frac{5}{14\beta_\pi (u\cdot p)} p^\alpha p^\beta \pi_\alpha^\gamma \pi_{\beta\gamma}  
+\frac{\tau_\pi}{3(u\cdot p)} p^\alpha p^\beta \pi_{\alpha\beta} \Theta      \nonumber \\
&&-\frac{6\tau_\pi}{5}p^\alpha \dot{u}^\beta \pi_{\alpha\beta}
 +\frac{(u\cdot p)}{70\beta_\pi}\pi^{\alpha\beta}\pi_{\alpha\beta}
 +\frac{\tau_\pi}{5}p^\alpha (\nabla^\beta\pi_{\alpha\beta})   \nonumber \\
&&-\frac{3\tau_\pi}{ (u\cdot p)^2} p^\alpha p^\beta p^\gamma \pi_{\alpha\beta} \dot{u}_\gamma
+ \frac{\tau_\pi}{2 (u\cdot p)^2} p^\alpha p^\beta p^\gamma (\nabla_\gamma\pi_{\alpha\beta})
\nonumber \\
&&-\frac{\beta + (u \cdot p)^{-1}}{4 (u \cdot p)^2 \beta_\pi} (p^\alpha p^\beta \pi_{\alpha\beta})^2\Bigg]; \label{Eq:visc_correction}
 \end{eqnarray}
 with, $\tau_\pi$ being the shear relaxation time, $\beta=1/T$, and $\tilde{f}_i^0 = 1 \pm f_i^0$ denoting the quantum statistical effects.

\subsubsection{Grad's ansatz}

In the Grad's 14-moment approximation, the viscous correction can be determined from a Taylor-like expansion in powers of momenta. Approximating the expansion up to second order in momenta, we get the function~\cite{Grad:1949zza,Israel:1979wp}
\begin{eqnarray}
    \delta f^G_i = f^0_i \tilde{f}_i^0 \left[\phi(x) + \phi_\alpha (x) p^\alpha + \phi_{\alpha\beta}(x) p^\alpha p^\beta \right];
\end{eqnarray}
where $\phi$s contain the information regarding dissipation and are independent of momenta. Casting the above expression in an orthogonal basis, we get
\begin{eqnarray}
     \delta f^G_i = f^0_i \tilde{f}_i^0 \left[\phi(x) + \phi_\alpha (x) p^{ \langle \alpha \rangle} + \phi_{\alpha\beta}(x) p^{ \langle \alpha} p^{\beta \rangle} \right]; 
\end{eqnarray}
where $p^{\langle \mu \rangle} \equiv \Delta _\nu^\mu p^\nu$. From the above expression, one identifies the $\phi$s to be bulk viscous pressure $\Pi$, particle four-current $n^\mu$, and shear stress tensor $\pi^{\mu\nu}$. 
In the absence of net conserved charges and for a conformal system, this approximation will give~\cite{Grad:1949zza,Israel:1979wp}
\begin{eqnarray}
    \delta f_i^G = \frac{f_i^0 \tilde{f}_i^0 \beta^2}{10 \beta_\pi} p^\alpha p^\beta \pi_{\alpha\beta}.
\end{eqnarray}
It must be noted that both the first order CE correction, $i.e.$, $\delta f^{(1)}$, and Grad's ansatz $\delta f^G$, lead to the the same shear-stress evolution equation, namely Eq.~\eqref{Eq:shear-evo-2}, with identical coefficients. 

%
\subsection{Bj\"orken Flow and Causal Hydrodynamics}
\label{sec:bjorken}

In our analysis, we employ the one-dimensional boost-invariant flow of Bj\"orken~\cite{Bjorken:1982qr}. The Bj\"orken's prescription is based on the assumption that matter formed in the heavy-ion collision expands only along the longitudinal beam ($z$) direction. To describe the dynamics of quark-gluon matter, it is convenient to adopt Milne coordinates $x^\mu =(\tau, x, y, \eta_s$) with the metric $g_{\mu\nu} = \textrm{diag}(+1, -1, -1, -\tau^2)$, where $\tau = \sqrt{t^2 - z^2}$ and $\eta_s = \textrm{arctanh}(z/t)$ are the proper-time and space-time rapidity, respectively. In these coordinates, the fluid four-velocity within Bj\"orken model becomes $u^\mu = (1, 0, 0, 0)$. 

Within Bj\"orken expansion, $\omega^{\mu\nu}=0$ and $\dot{u}^\mu=0$, and the scalar expansion rate is $\Theta = 1/\tau$. 
The shear-stress tensor is symmetric, traceless $\pi_\alpha^\alpha=0$ and orthogonal to four-velocity, $u^\mu \pi_{\mu\nu} =0$. Hence, $\pi^{\mu\nu}$ can be parameterized as
\begin{eqnarray}   \label{Eq:Shear}
    \pi^{\mu\nu} = \textrm{diag}\left(0, \frac{\pi}{2}, \frac{\pi}{2}, -\frac{\pi}{\tau^2} \right);
\end{eqnarray}
where $\pi \equiv - \tau^2 \pi^{\eta_s \eta_s}$. The evolution equations for energy density $\epsilon$ and shear stress $\pi$ (Eq.~\eqref{Eq:shear-evo-2}) within Bj\"orken geometry are given by~\cite{Jaiswal:2013fc}
\begin{eqnarray}   
    \frac{d\epsilon}{d\tau} &=& - \frac{1}{\tau}(\epsilon + P - \pi),  \label{Eq:epsilon-evolution}\\
    \frac{d\pi}{d\tau} &=& -\frac{\pi}{\tau_\pi} + \frac{4\beta_\pi}{3\tau} - \frac{\lambda \pi}{\tau};   \label{Eq:pi-evolution}
\end{eqnarray}
where, the transport coefficients are given by $\tau_\pi = \eta /\beta_\pi$ and $\lambda =38/21$. These equations are closed by providing the conformal equation of state (EoS) for relativistic non-interacting massless quarks and gluons:
\begin{eqnarray}
    \epsilon = 3P = 3 \left[ 2(N_c^2 -1) + \frac{7}{2}N_c N_f\right] \frac{\pi^2}{90} T^4;
\end{eqnarray}
where $N_c=3$ and $N_f =2$ in our analysis. In the absence of viscosity, the evolution equation \eqref{Eq:epsilon-evolution} gives the wellknown ideal Bj\"orken solution $T(\tau) = (\tau_0/\tau)^{1/3}T_0$; where, $T_0$ and $\tau_0$ are the initial temperature and proper time, respectively.

\subsection{Momentum dependence of $\delta f$} \label{sec:momentum-dep-deltaf}

\begin{figure*} 
   \centering
     \subfigure
     []{
    \includegraphics[width=0.49\textwidth, height=7.0cm]{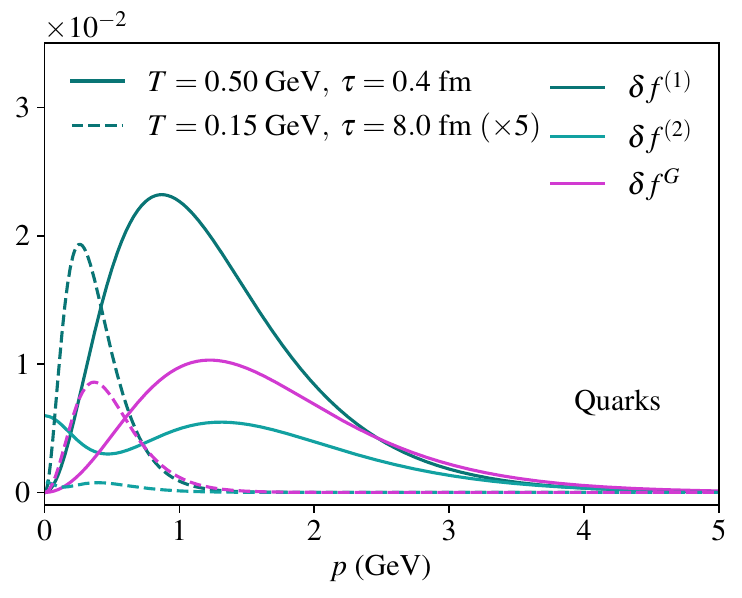}\label{fig:deltaf-1}} 
    \subfigure[]{\includegraphics[width=0.49\textwidth]{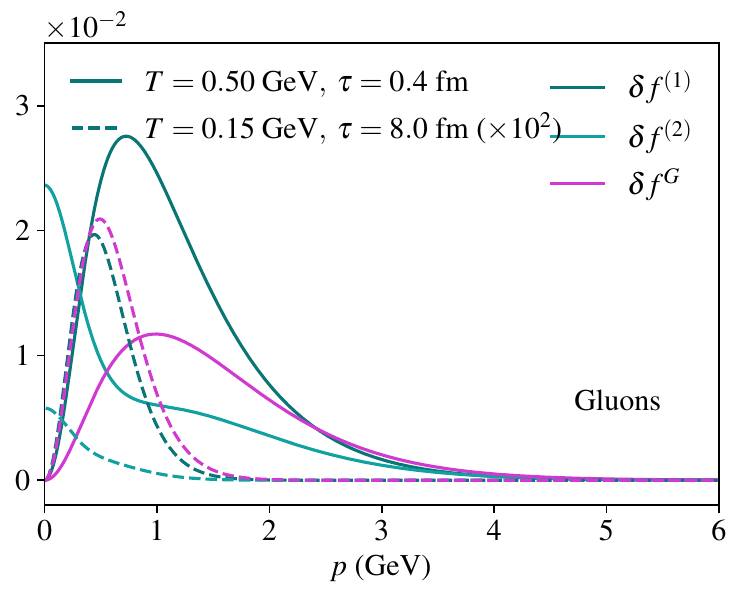}\label{fig:deltaf-2}}
  \caption{The momentum dependence of viscous corrections for (a) quarks and (b) gluons considering quantum statistics and thermal masses, for representative early and late times of Bj\"orken evolution.
  }
  \label{fig:deltaf}
  \end{figure*}
The structure of viscous corrections to the phase-space distribution function plays a central role in determining the observable dependence of non-equilibrium effects. In order to study the behaviour of viscous corrections, we adopt the Minkowski coordinates and cast the non-equilibrium parts of particle distribution functions within Bj\"orken flow considering the local rest frame (LRF), $u^\mu = (1, \vec{0})$. Using Eq.~\eqref{Eq:Shear}, we write the viscous corrections from both CE-like expansion and Grad's method as follows:
\begin{eqnarray}
    \delta f_i^{CE} &=& \frac{f_i^0 \bar{f}_i^0 \beta}{\beta_\pi}\frac{s}{E \tau} \left(\frac{\eta}{s} \right) \left( \frac{p^2}{3} - {p_z^2} \right) \nonumber\\
    &&- \frac{f_i^0 \bar{f}_i^0 \beta}{\beta_\pi} \Bigg[ 
    \frac{10}{21} \frac{s^2}{\beta_\pi E \tau^2} \left(\frac{\eta}{s}\right)^2 \left( \frac{p^2}{3} + p_z^2 \right) \nonumber \\
    &&+ \frac{2}{3} \frac{\tau_\pi s}{E \tau^2} \left(\frac{\eta}{s} \right) \left( \frac{p^2}{3} - p_z^2\right) 
    + \frac{4}{105} \frac{E s^2}{\beta_\pi \tau^2} \left(\frac{\eta}{s}\right)^2 \nonumber\\
    &&- \frac{4}{15} \frac{\tau_\pi E s}{\tau^2} \left(\frac{\eta}{s}\right) + \frac{4}{3} \frac{\tau_\pi s}{E\tau^2}\left( \frac{\eta}{s}\right) p_z^2 \nonumber\\
    &&- \frac{\beta + E^{-1}}{E^2 \beta_\pi} \frac{s^2}{\tau^2} \left(\frac{\eta}{s}\right)^2 \left(\frac{p^2}{3} - p_z^2 \right)^2 
    \Bigg],  \label{Eq:deltaf_CE}\\
    \delta f_i^G &=& \frac{f^0_i \bar{f}_i^0}{ T^3 \tau} \left(\frac{\eta}{s} \right) \left( \frac{p^2}{3} - {p_z^2} \right).  \label{Eq:deltaf_G}
\end{eqnarray}
We note that, in the above equations, the Navier-Stokes value for shear stress, $\pi = 4\eta/(3\tau)$ is used.
\par
In Figs.~\ref{fig:deltaf-1} and \ref{fig:deltaf-2}, we plot the momentum dependence of viscous corrections for quarks and gluons respectively, including the quantum statistics and thermal masses of the particles, for representative early and late-time evolution conditions. We take the following expressions for the thermal masses for quarks and gluons~\cite{GolamMustafa:1997id} $m^2_q(T)=g^2T^2/6,\,m_g^2(T)=(1+N_f/6)g^2T^2/3$, respectively and we fix $\eta/s = 1/4\pi$ and $p_z =0$ for this analysis. A clear hierarchy is observed in the momentum dependence of the different contributions. At early time, it is seen that the first order CE correction $\delta f^{(1)}$ increases sharply in the thermal momentum region $p \sim T$, peaks in the intermediate momentum range and then decreases with $p$. Importantly, we observe that the correction due to the second order terms in CE expansion, $\delta f^{(2)}$ remains sub-leading and decreases with $p$. However, at very low momenta, $\delta f^{(2)}$ is observed to be larger, especially for the gluons (Fig.~\ref{fig:deltaf-2}), which shows stronger low momentum contribution compared to quarks. The Grad's ansatz $\delta f^G$ has a broader momentum dependence and is seen to be smaller compared to $\delta f^{(1)}$ in the region $p< 2.5$ GeV. While at large momenta, $\delta f^G$ becomes dominant owing to the quadratic momentum dependence of the correction, Eq.~\eqref{Eq:deltaf_G}. At late time, the contributions from different viscous corrections become weaker, while they show qualitatively similar behaviour. 
\par
We note that understanding the momentum dependence of $\delta f$ is crucial to study the observable dependence of these viscous corrections. This hierarchy in the momentum structure of $\delta f$ forms the basis in determining the observable sensitivity of heavy-quark transport and dilepton production to non-equilibrium dynamics in the following sections.

\section{Heavy Quark Transport}
\label{sec:heavy-quark}

\subsection{Drag and diffusion coefficients}

Heavy quarks (HQs)  are produced during the initial stage of the collision via hard scatterings, and they survive throughout the expansion of QGP. The drag and diffusion coefficients carry the interactions of HQs with the constituent quarks and gluons. Owing to their large mass ($ m_c \gg T$), HQs interact with the medium predominantly through multiple soft scatterings, leading to a stochastic evolution in momentum space. The propagation of heavy quarks in the QGP medium can be described using the Boltzmann transport equation,
\begin{equation}
\frac{\partial f(\mathbf{p})}{\partial t} = \mathcal{C}[f];
\end{equation}
where, $f(\bf p)$ is the phase-space distribution function of HQ and $\mathcal{C}[f]$ denotes the collision integral, which encodes the rate of change of $f(\bf p)$ due to interactions with the quarks and gluons in the medium. For binary interactions, the collision term may be written as
\begin{equation}
\mathcal{C}[f] = \int d^3{\bf k} \left[ w(\mathbf{p}+\mathbf{k},\mathbf{k}) f(\mathbf{p}+\mathbf{k}) - w(\mathbf{p},\mathbf{k}) f(\mathbf{p}) \right]. \label{Eq:collision-term-1}
\end{equation}
Here, $w(\mathbf{p},\mathbf{k})$ corresponds to the rate at which the heavy quark momentum changes from ${\bf p}$ to ${\bf p-k}$. 
A more microscopic representation of the collision term, based on $2 \to 2$ scattering processes, is given by~\cite{Svetitsky:1987gq}
\begin{align}
\mathcal{C}[f] &=
\frac{1}{2E_p}
\int \frac{d^3 {\bf q}}{(2\pi)^3\,2E_q}
\int \frac{d^3 {\bf q'}}{(2\pi)^3\,2E_{q'}}
\int \frac{d^3 {\bf p'}}{(2\pi)^3\,2E_{p'}}  \nonumber \\
&\quad \times \frac{1}{\gamma_c}
\sum |\mathscr{M}|^2 \,
(2\pi)^4 \delta^{4}(p + q - p' - q') \nonumber \\
&\quad \times \Big[
f(\mathbf{p}')\, g(\mathbf{q}')\, \tilde{g}(\mathbf{q})
- f(\mathbf{p})\, g(\mathbf{q})\, \tilde{g}(\mathbf{q}')
\Big];   \label{Eq:collision-term2}
\end{align}
where, $E_p = \sqrt{{\bf p}^2 + m_c^2}$ is the energy of the HQ (charm quark) and $E_q = \sqrt{{\bf q}^2 + m_{q,g}^2}$ denote the energy of quarks/gluons in the medium. Also, ${\bf p'} = {\bf p-k}$ and ${\bf q'} = {\bf q + k}$ are the momenta of the HQ and medium particles after scattering. Here, $g(\mathbf{q})$ denotes the distribution function of constituent quarks/gluons and $\tilde{g}(\mathbf{q}) = 1 \pm g(\mathbf{q})$ accounts for quantum statistical effects. The squared matrix element $\sum |\mathscr{M}|^2$ encodes the interaction dynamics and in this analysis, we use the $\sum |\mathscr{M}|^2$ for quark and gluon scatterings as defined in Ref.~\cite{Svetitsky:1987gq}. In the above equation, $\gamma_c = 6$ is the degeneracy factor of HQs. 

To simplify the collision kernel, we employ the Landau approximation~\cite{Landau:1937vum}, which assumes the soft momentum transfer in the particle scattering $|\mathbf{k}| \ll |\mathbf{p}|$. In Eq.~\eqref{Eq:collision-term-1}, $\mathcal{C}[f]$ is now expanded in powers of $\mathbf{k}$ up to second order, which results in the Fokker-Planck equation,
\begin{equation}
\frac{\partial f}{\partial t}
=
\frac{\partial}{\partial p_i}
\left[
A_i(\mathbf{p}) f
+
\frac{\partial}{\partial p_j}
\left[ B_{ij}(\mathbf{p}) f \right]
\right],
\end{equation}
where $A_i$ is the drag force and $B_{ij}$ is the momentum diffusion of HQs in QGP medium with $i,j=1,2,3$ and these transport coefficients are defined as
\begin{align}
A_i(\mathbf{p}) &= \int d^3{\bf k}\, w(\mathbf{p},\mathbf{k})\, k_i, \\
B_{ij}(\mathbf{p}) &= \frac{1}{2} \int d^3{\bf k}\, w(\mathbf{p},\mathbf{k})\, k_i k_j.
\end{align}
Using Eqs.~\eqref{Eq:collision-term-1} and \eqref{Eq:collision-term2}, these coefficients can be expressed in terms of the underlying microscopic scattering processes as~\cite{Svetitsky:1987gq},
\begin{align}\label{Ai}
A_i(\mathbf{p})
&= \frac{1}{2E_p}
\int \frac{d^3 {\bf q}}{(2\pi)^3\,2E_q}
\int \frac{d^3 {\bf q'}}{(2\pi)^3\,2E_{q'}}
\int \frac{d^3 {\bf p'}}{(2\pi)^3\,2E_{p'}}  \nonumber \\
&\quad \times \frac{1}{\gamma_c}
\sum |\mathscr{M}|^2 \,
(2\pi)^4 \delta^{4}(p + q - p' - q') \nonumber \\
&\quad \times g(\mathbf{q})\, \tilde{g}(\mathbf{q}')
\, (p - p')_i
\;\equiv\; \langle\!\langle \Delta p_i \rangle\!\rangle ,
\end{align}
and
\begin{equation}
B_{ij}(\mathbf{p})
= \frac{1}{2}
\left\langle\!\left\langle
\Delta p_i \Delta p_j
\right\rangle\!\right\rangle ;
\end{equation}
where, $\Delta p_i = (p - p')_i$ denotes the momentum transfer in a single scattering. The drag and diffusion coefficients of HQs represent the thermal average of the momentum transfer and its square respectively, when interacting with medium particles. 

Since $A_i$ and $B_{ij}$ are dependent only on ${\bf p}$, they can be decomposed as
\begin{eqnarray*}
A_i(\mathbf{p}) &=& p_i\, A(p^2), \\
B_{ij}(\mathbf{p}) &=&
\left(\delta_{ij} - \frac{p_i p_j}{p^2}\right) B_0(p^2)
+ \frac{p_i p_j}{p^2} B_1(p^2),
\end{eqnarray*}
where $B_0$ and $B_1$ correspond to transverse and longitudinal diffusion, respectively. Here, the scalar functions are given by
\begin{eqnarray}
A &=& \langle\!\langle 1 \rangle\!\rangle - \frac{\langle\!\langle \mathbf{p}\cdot \mathbf{p}' \rangle\!\rangle}{p^2},  \label{Eq:drag} \\
B_0 &=& \frac{1}{4} \left[
\langle\!\langle p'^2 \rangle\!\rangle
- \frac{\langle\!\langle (\mathbf{p}' \cdot \mathbf{p})^2 \rangle\!\rangle}{p^2}
\right],    \\
B_1 &=& \frac{1}{2} \left[
\frac{\langle\!\langle (\mathbf{p}' \cdot \mathbf{p})^2 \rangle\!\rangle}{p^2}
- 2 \langle\!\langle \mathbf{p}' \cdot \mathbf{p} \rangle\!\rangle
+ p^2 \langle\!\langle 1 \rangle\!\rangle
\right].
\end{eqnarray}
The phase-space integrals can be conveniently evaluated in the center-of-mass frame, leading to~\cite{GolamMustafa:1997id}
\begin{align}
\langle\!\langle \psi({\bf p'}) \rangle\!\rangle
&= \frac{1}{512\pi^4 \gamma_c} \frac{1}{E_p}
\int_0^\infty \frac{|{\bf q}|^2\,d|{\bf q}|}{E_q}\, 
\int_{-1}^{1} d(\cos\chi) \nonumber \\
&\quad \times 
\frac{\left[(s + m_c^2 - m_{q,g}^2)^2 - 4 s m_c^2 \right]^{1/2}}{s}
\, g(E_q) \nonumber \\
&\quad \times \int_{-1}^{1} d(\cos\theta_{\mathrm{c.m.}})
\sum |\mathscr{M}|^2
\int_0^{2\pi} d\phi_{\mathrm{c.m.}} \nonumber \\
&\quad \times e^{\beta E_{q'}}\, g(E_{q'})\, \psi({\bf p'}) \, . \label{Eq:hq-integral}
\end{align}
Here, $s=(p+q)^2$ is the Mandelstam variable, $\theta_{\mathrm{c.m.}}$ and $\phi_{\mathrm{c.m.}}$ denote the scattering angles in the center-of-mass frame.

Quantum statistical effects can lead to infrared divergences in the limit $q \to 0$, which are regulated by introducing thermal masses. In the present calculation, medium screening is incorporated through the Debye mass, $\mu_D = \sqrt{4\pi \alpha_s}\, T$, with $\alpha_s = 0.3$. Also, we take 
the charm quark mass and degeneracy to be $m_c=1.5$ GeV and $\gamma_c=6$.  

Further, we look into the spatial diffusion coefficient $D_s$, which measures the HQ interactions with medium~\cite{He:2022ywp}
\begin{equation}
 D_s = \frac{m_c}{TA(p \to 0)} ,
\end{equation}
with second order viscous correction. The spatial diffusion coefficient has been estimated via lattice QCD calculations as well~\cite{HotQCD:2025fbd} and is directly related to the thermalization time 
$\tau_{th} = m_c D_s/T$.  

\subsection{Viscous corrections to heavy-quark transport}
\label{subsec:HQ_viscous}

The interaction of heavy quarks with the medium is governed by the distribution function of the bath particles entering the collision integral. In a viscous medium, this distribution deviates from equilibrium, $g_i(\mathbf{q}) = g^0_i(\mathbf{q}) + \delta g_i(\mathbf{q})$,
where $g^0_i$ is the equilibrium distribution and $\delta g_i$ denotes the non-equilibrium correction. The explicit form of $\delta g_i$ depends on the choice of non-equilibrium ansatz. In the Chapman-Enskog (CE) approach, the correction up to second order is 
$\delta g_i^{CE} = \delta g_i^{(1)} + \delta g_i^{(2)}$, 
where $\delta g_i^{(1)}$ and $\delta g_i^{(2)}$ denote the first- and second-order contributions, respectively, given by Eq.~\eqref{Eq:deltaf_CE}. 
Substituting this into the microscopic expression for the transport coefficients, Eq.~\eqref{Ai}, the drag coefficient can be decomposed as     
\begin{equation}
A_i = A_i^{id} + \delta A_i;  \label{Eq:A_sum}
\end{equation}
where, $A_i^{id}$ corresponds to the equilibrium contribution and $\delta A_i$ encodes viscous corrections. 
In the present work, we consider upto second order in deviations from equilibrium- $\delta g^{(2)}$ and neglect higher-order products of $\delta g$ to obtain
\begin{align}
\delta A_i(\mathbf{p})
&= \frac{1}{2E_p}
\int \frac{d^3 \mathbf{q}}{(2\pi)^3\,2E_q}
\int \frac{d^3 \mathbf{q}'}{(2\pi)^3\,2E_{q'}}
\int \frac{d^3 \mathbf{p}'}{(2\pi)^3\,2E_{p'}} \nonumber\\
&\quad \times \frac{1}{\gamma_c}
\sum |\mathscr{M}|^2 \,
(2\pi)^4 \delta^{(4)}(p + q - p' - q') \nonumber \\
&\quad \times \Big[
\{\delta g^{(1)}(\mathbf{q})+\delta g^{(2)}(\mathbf{q})\}\, \tilde{g}^0(\mathbf{q}') \, \nonumber\\ 
& \qquad \,+ g^0(\mathbf{q})
\{\delta \tilde{g}^{(1)}(\mathbf{q}')+\delta \tilde{g}^{(2)}(\mathbf{q}')\} \nonumber\\ 
& \qquad \,+\delta g^{(1)}(\mathbf{q})\delta \tilde{g}^{(1)}(\mathbf{q}')
\Big]\, (p - p')_i . \label{Eq:drag-visc-part}
\end{align}
Note that while using the Grad's ansatz $\delta g^{G}$, the correction enters linearly through the modification of the medium distribution with the term 
$\Big[\delta g(\mathbf{q})\, \tilde{g}^0(\mathbf{q}')
+ g^0(\mathbf{q})\, \delta \tilde{g}(\mathbf{q}')
\Big]$ in the above equation, leading to a different redistribution of viscous corrections in momentum space. Also, the effect of viscosity on the diffusion coefficients $B_0$ and $B_1$ can be treated in a similar manner as described in Eq.~\eqref{Eq:A_sum}.

\subsection{Momentum space structure}
For the purpose of analysis of structure of the momentum integral in presence of non-equilibrium corrections, it is convenient to express the drag coefficient schematically as 
\begin{equation}
A(\mathbf{p}) = \int d |\mathbf{q}|\,\mathbf{|q|}^2 \; W_A(\mathbf{p},\mathbf{q}),
\end{equation}
where $W(\mathbf{p}, \mathbf{q})$ encodes the scattering kernel, including the matrix element and final-state phase space. The viscous correction to drag then takes the form
\begin{equation}
\delta A(\mathbf{p}) = \int d |\mathbf{q}|\,\mathbf{|q|}^2 \; \delta W_A(\mathbf{p},\mathbf{q}).
\end{equation}
Here, $W_A(\mathbf{p}, \mathbf{q})$ and $\delta W_A(\mathbf{p}, \mathbf{q})$ are calculated using the Eqs.~\eqref{Eq:drag}, \eqref{Eq:hq-integral}, with the product of viscous corrections in $\delta W_A(\mathbf{p}, \mathbf{q})$ taken as in Eq.~\eqref{Eq:drag-visc-part}. This representation makes it explicit that the viscous modification of heavy-quark transport is governed by the interplay between the momentum dependence of $\delta g(\mathbf{q})$ and the structure of the scattering kernel $W(\mathbf{p}, \mathbf{q})$ and we proceed to determine the relative importance of different viscous corrections. Here, we note that the kernels for diffusion coefficients $B_0$ and $B_1 - B_0$ can also be determined in a similar manner. 
\par
\begin{figure*} 
   \centering
     \subfigure
     []{
    \includegraphics[width=0.49\textwidth, height=6.7cm]{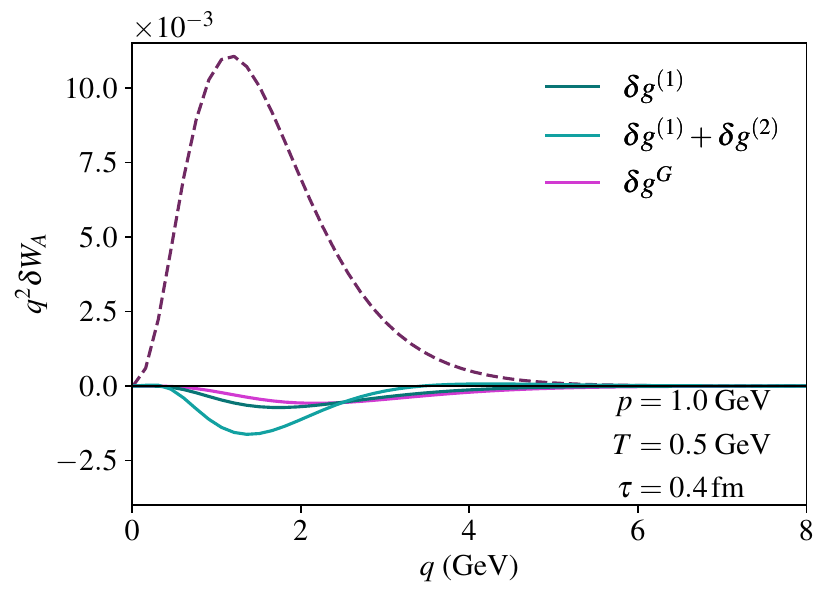}\label{fig:A-kernel-initial}} 
    \subfigure[]{\includegraphics[width=0.49\textwidth]{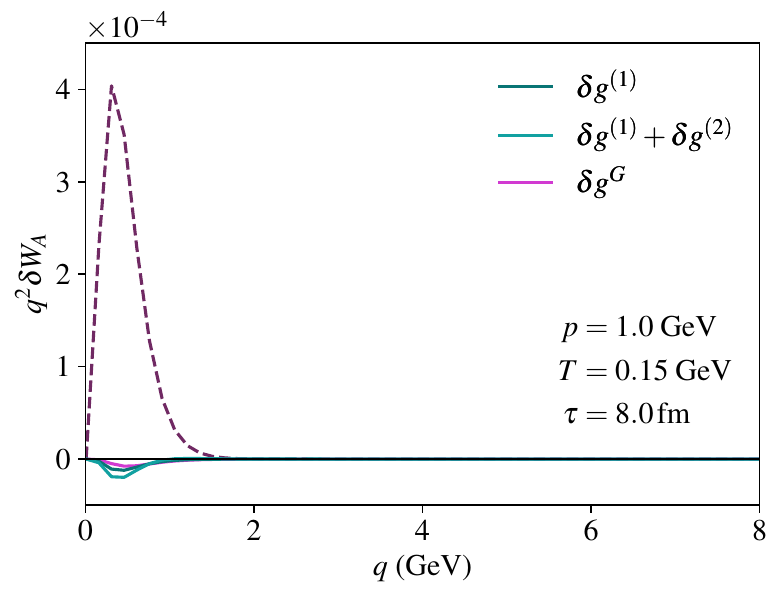}\label{fig:A-kernel-final}}
  \caption{Individual contributions of different viscous corrections to the kernel of drag force for an (a) early and (b) late time evolution of collision, fixing $p =1$ GeV. Drag kernel in the absence of viscous corrections (dashed curves) is also shown. 
  }
  \label{fig:A-kernel}
  \end{figure*}

In Fig.~\ref{fig:A-kernel-initial}, we plot the viscous contributions to the kernel of drag force, $q^2 \delta W_A$ as a function of momentum $q$ of interacting particles at an early time (high temperature and strong gradients), fixing the HQ momentum to be $p=1$ GeV. For comparison, we also plot the ideal contribution $q^2 W_A^{id}$ (dashed curves). The drag kernel is positive in the ideal case and has a peak around $q \sim 1$ GeV. We see that the contribution of viscous corrections begins from zero and becomes negative for small $q$, indicating that the drag kernel is suppressed in this region. The modification due to $\delta g^{(1)}$ increases around $q \sim 2$ GeV, and then decreases with increment in $q$. When the second-order terms are included, the non-equilibrium contribution further increases at small $q$ leading to maximum suppression of drag coefficient in the thermal momentum regime. We find that the modification due to $\delta g^{(2)}$ term reduces the total CE contribution by partially canceling the effect from $\delta g^{(1)}$ with increment in $q$ and the net correction becomes slightly positive. 
It can be seen that for $q>3$ GeV, the contribution from $\delta g^{CE}=\delta g^{(1)} + \delta g^{(2)}$ becomes moderate and that from $\delta g^G$ becomes dominant. Also, with increasing $q$, the effect of viscous corrections on the kernel becomes negligible. At a later time of expansion (low temperature), from Fig.~\ref{fig:A-kernel-final}, we see that the contributions arising from different corrections remain qualitatively similar, while their magnitudes reduce by two orders. Also, the effect of corrections limits to the low momentum region $q < 2$ GeV. 
\par
\begin{figure*} 
   \centering
     \subfigure
     []{
    \includegraphics[width=0.49\textwidth]{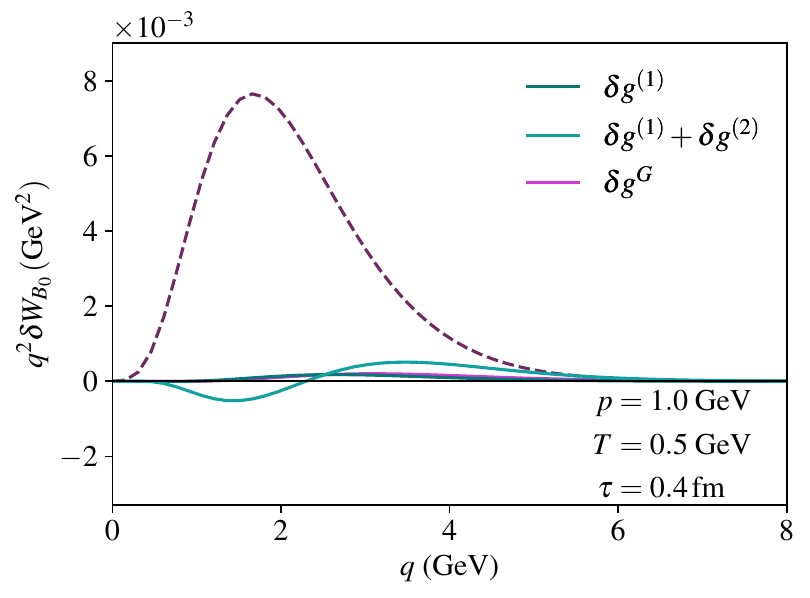}\label{fig:B0-kernel-initial}} 
    \subfigure[]{\includegraphics[width=0.49\textwidth, height=6.5cm]{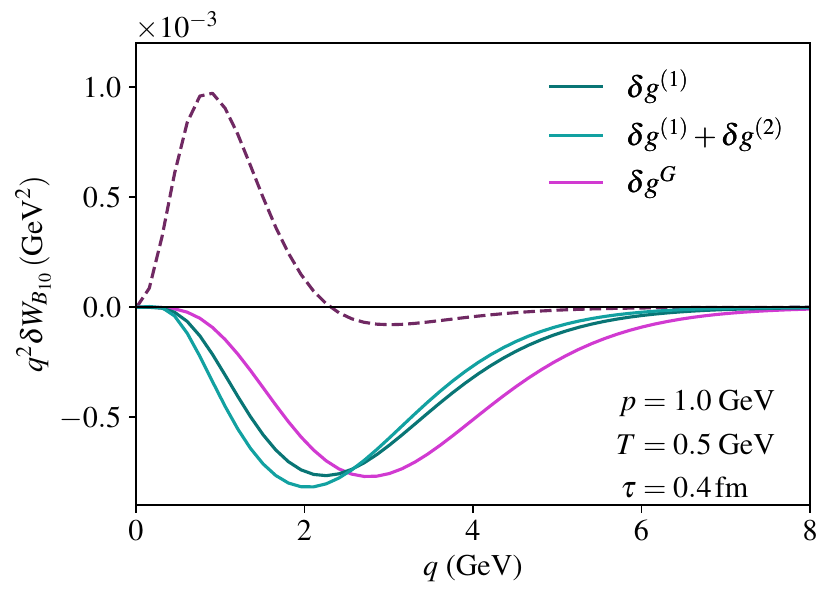}\label{fig:B10-kernel-initial}}
  \caption{Individual contributions of different viscous corrections to the kernel of coefficients (a) $B_0$ and (b) $B_1 - B_0$ for an early time evolution, fixing $p =1$ GeV. Corresponding kernels in the absence of viscous corrections are also shown.}
  \label{fig:B-kernel}
  \end{figure*}

Similarly, we show the kernels $q^2 \delta W_{B_0}$ and $q^2 \delta W_{B_{10}}$ corresponding to the transverse diffusion coefficient $B_0$ and measure of momentum diffusion anisotropy $B_1 - B_0$, respectively, in Fig.~\ref{fig:B-kernel}, by fixing $p=1$ GeV. The contribution from the ideal case is also displayed for comparison. In the case of transverse diffusion kernel (Fig.~\ref{fig:B0-kernel-initial}), we observe that the correction due to $\delta g^{(1)}$ remains positive throughout the momentum range. When $\delta g^{(2)}$ contribution is included, modification due to the total CE correction is increased. The correction from $\delta g^{(1)} + \delta g^{(2)}$ is seen to be dominant throughout the range of momentum $q$, unlike the structure of drag kernel. Also, the modification arising from $\delta g^G$ to the $B_0$ kernel is slightly large than that from $\delta g^{(1)}$ for $q>3$ GeV. Further, as seen for the drag kernel, the contributions of different viscous corrections decrease with an increase in $q$.
Now, in Fig.~\ref{fig:B10-kernel-initial}, we observe that the contribution from $\delta g^{CE}$ appears large for the quantity $B_1 - B_0$ in the momentum range $q < 3$ GeV. For $q>3$ GeV, the second-order terms partially cancel the corrections from $\delta g^{(1)}$ thus reducing the net CE contribution. The effect of $\delta g^G$ on $B_1 - B_0$ is found to be dominant for the region $q>3$ GeV. Further, unlike the kernels of $A$ and $B_0$, here we see a distinct hierarchy between the different viscous contributions. Also, we observe that the impact of viscous corrections remain for a wider range of $q$. The kernels at late times of evolution are not shown since they behave qualitatively similar with smaller magnitudes, as seen before. 
\par
From the transport kernels depicted here, we demonstrate that the viscous effects are strong at early evolution times, at which the medium is hot and the viscous gradients are large. Due to cooling and longitudinal expansion of the medium, the viscous effects diminish at late times. This feature is as observed from the momentum dependence of viscous corrections. Further, we find that these viscous effects majorly alter the momentum structure of the coefficients at small values of $q$, where the integral has maximum contribution. The momentum structure of transport coefficients plotted here demonstrates that the viscous modifications of these coefficients are governed not only by the magnitude of the non-equilibrium correction itself, but more importantly by its momentum overlap with the transport kernel. We also note that neglecting quantum effects are not altering the results much and coefficients show same qualitative behaviour in the Maxwell-Boltzmann limit.

\section{
Thermal Dilepton production}
\label{sec:dilepton}
\subsection{Second order viscous corrections to dilepton rate}

Dileptons are created from all the stages of heavy-ion collisions through various mechanisms and are considered to be one of the promising probes of QGP. Here, we consider the dominant processes contributing to thermal dilepton production in the QGP and hadronic phases - $q\bar{q}$-annihilation, $q\bar{q} \rightarrow \gamma^\ast \rightarrow l^{+}l^{-}$ and $\pi^{+}\pi^{-}$-annihilation, $\pi^{+}\pi^{-} \rightarrow \rho^{0} \rightarrow l^{+}l^{-}$, respectively. 
From relativistic kinetic theory, the rate of thermal dilepton production for these processes is given by~\cite{Vogt:2007zz}
\begin{eqnarray}
 \frac{dR}{d^4p} &=& \int \frac{d^3 {\bf p}_1}{(2 \pi)^3} \frac{d^3 {\bf p}_2}{(2\pi)^3}
 f({\bf p}_1) f({\bf p}_2) \frac{M^2 \gamma^2 \sigma(M^2)}{2 E_1 E_2} \nonumber\\
 &&\quad \times\delta^4(p-p_1-p_2); 
\end{eqnarray}
where, $p_{1,2} = (E_{1,2}, {\bf p}_{1,2})$ denote the four-momenta of the incoming particles having masses $m_{1,2}$, and $p = (E = E_1 + E_2, {\bf p} = {\bf p}_1 + {\bf p}_2)$ represent the four-momentum of the dilepton, with $M^2 = (E_1 + E_1)^2 - ({\bf p}_1 + {\bf p}_2)^2$ being the square of invariant mass. In the above equation, $\gamma$ is the degeneracy factor for interacting particle and $\sigma(M^2)$ is the cross-section for the process. With $N_f =2$ and $N_c = 3$, we have $M^2 \gamma^2 \sigma_q(M^2) = 80 \pi \alpha_e^2/9$ for the $q \bar{q}$-annihilation process, where $\alpha_e=1/137$ is the electromagnetic coupling constant. Also, we get $M^2 \gamma^2 \sigma_\pi(M^2) = (4\pi/3)\alpha_e^2 |F_\pi(M^2)|^2$~\cite{Vogt:2007zz} for the $\pi^{+}\pi^{-}$-annihilation process. The term $|F_\pi(M^2)|^2 = m_\rho^4/[(m_\rho^2 - M^2)^2 + m_\rho^2 \Gamma_\rho^2]$ is the electromagnetic pion form factor with $m_\rho = 775$ MeV and $\Gamma_\rho =149$ MeV being the mass and decay width of $\rho(770)$ meson respectively~\cite{Song:2010fk}.
\par
The presence of viscosity modifies the particle distribution functions as $f_i ({\bf p}) = f_i^0 ({\bf p}) + \delta f_i ({\bf p})$. In this work, we consider different forms of viscous correction $\delta f_i$, as defined in Sec.~\ref{sec:visc-correction}. Below, we present the calculation of thermal dilepton rate using the $\delta f$ up to second-order in momenta obtained from the Chapman-Enskog like expansion (Eq.~\eqref{Eq:visc_correction}), {\it i.e.,} $\delta f^{CE}_i = \delta f_i^{(1)} + \delta f_i^{(2)}$.  
Keeping the terms only up to fourth-order in momenta, we can write the dilepton rate as
\begin{equation}
 \frac{dR}{d^4p} = \frac{dR^{id}}{d^4p} + \frac{dR^{(1)}}{d^4p} 
                      + \frac{dR^{(2)}}{d^4p};   \label{Eq:total-rate-expression}
\end{equation}
with the ideal part of the dilepton rate is given by~\cite{Vogt:2007zz}
\begin{eqnarray}
 \frac{dR^{id}}{d^4p} = \frac{\mathscr{R} }{2}  e^{-(u \cdot p)/T};  \label{Eq:dil-rate-ideal}
\end{eqnarray}
where, $\displaystyle \mathscr{R} =  \sum_{i=q,\pi} M^2 \gamma^2 \sigma_i(M^2) / (2 \pi)^5$.
The contribution to the dilepton rate from the first-order CE viscous correction can be obtained as~\cite{Naik:2021yph,Naik:2021yue,Naik:2025pjt}
\begin{eqnarray}
     \frac{dR^{(1)}}{d^4p} &=& \frac{dR^{id}}{d^4p}\frac{\beta}{\beta_\pi}  \Bigg(  \frac{\mathscr{E} (u\cdot p) }{2} \left[2\mathscr{E}^2 - 3M^2\right] +\frac{3}{4}M^4 \mathscr{Y} \Bigg) \nonumber \\
     && \times \frac{p^\mu p^\nu \pi_{\mu\nu}}{2\mathscr{E}^{5}}; \label{Eq:dil-rate-Iorder}
\end{eqnarray}
where, 
\begin{eqnarray*}
\mathscr{E}^2 &=& (u\cdot p)^2-M^2 \,\,\, \textrm{and}\quad 
\mathscr{Y} = \ln\left[\frac{(u\cdot p)+\mathscr{E}}{(u\cdot p)-\mathscr{E}}\right].
\end{eqnarray*}
Viscous contribution to the rate due to second-order CE terms 
is calculated as follows. We adopt the Minkowski metric $\eta_{\mu\nu} = \textrm{diag}(+1, -1, -1, -1)$ in the following calculation. We decompose the dilepton rate as distinct contributions arising from the non-vanishing terms in $\delta f^{(2)}$ as
\begin{eqnarray}
    \frac{dR^{(2)}}{d^4p} &=& \frac{\beta}{\beta_\pi} \Bigg\lbrace 
    \frac{5}{7 \beta_\pi} I^{\alpha\beta} \pi_\alpha^\gamma \pi_{\beta\gamma} 
    - \frac{2\tau_\pi}{3} I^{\alpha\beta} \pi_{\alpha\beta} \Theta  \nonumber \\
    && - \frac{E}{70 \beta_\pi} \frac{dR_0}{d^4 p} \pi_{\alpha\beta} \pi^{\alpha\beta}
    - \frac{2\tau_\pi}{5} I^\alpha \nabla^\beta \pi_{\alpha\beta} 
     \nonumber \\
    && - {\tau_\pi} I^{\alpha\beta\gamma} \nabla_\gamma \pi_{\alpha\beta} + \frac{1}{2\beta_\pi} I^{\alpha\beta\gamma\delta} \pi_{\alpha\beta} \pi_{\gamma\delta}
    \Bigg\rbrace.  \label{Eq:dil-decompose}
\end{eqnarray}
In the above equation, we have represented
\begin{eqnarray}
I^\alpha &=& \int d{\bf \Gamma} \frac{\mathscr{R}}{2 E_1 E_2} \,p_2^\alpha\, \delta(E-E_1-E_2), \nonumber \\
 I^{\alpha\beta} &=& \int d{\bf \Gamma} \frac{\mathscr{R}}{2 E_1 E_2}\, \frac{p_2^\alpha p_2^\beta}{u \cdot p_2}\, \delta(E - E_1-E_2), \nonumber \\
 I^{\alpha\beta\gamma} &=& \int d{\bf \Gamma} \frac{\mathscr{R}}{2 E_1 E_2}\, \frac{p_2^\alpha p_2^\beta p_2^\gamma}{(u \cdot p_2)^2} \,\delta(E - E_1-E_2), \nonumber \\
 I^{\alpha\beta\gamma\kappa} &=& \int d{\bf \Gamma} \frac{\mathscr{R}}{2 E_1 E_2}\,  p_2^\alpha p_2^\beta p_2^\gamma p_2^\kappa\, \frac{\beta + (u \cdot p_2)^{-1}}{(u \cdot p_2)^2} \nonumber\\
 &&\quad \times  \delta(E - E_1-E_2), \label{Eq:tensor-I}
\end{eqnarray}
with 
\begin{eqnarray}
  d{\bf \Gamma} \equiv  \frac{d^3 {\bf p}_2}{2\pi}
 f_i^0({\bf p}_1) f_i^0({\bf p}_2). \nonumber
\end{eqnarray}

To evaluate the above integrals, we write the tensors constructed using the metric $\eta^{\alpha\beta}$, four-velocity $u^\mu$ and four-momentum $p^\mu$. Considering the local rest frame of the fluid $i.e.,$ $u^\mu = (1, \vec{0}$), the tensors can be written as
\begin{eqnarray}
    I^i &=& a_0 p^i  \nonumber \\
    I^{ij} &=& -b_0 \delta^{ij} + b_1 p^i p^j  \nonumber \\
    I^{ijk} &=& -c_0 (\delta^{ij} p^k + \delta^{ik} p^j + \delta^{kj} p^i) + c_1 p^i p^j p^k \nonumber \\
    I^{ijkl} &=& d_0 (\delta^{ij} \delta^{kl} + \delta^{ik} \delta^{jl} + \delta^{il}\delta^{jk}) + d_1 p^i p^j p^k p^l - d_2 (\delta^{ij} p^k p^l \nonumber \\
    &&  + \delta^{kl} p^i p^j + \delta^{ik} p^j p^l + \delta^{il} p^j p^l + \delta^{jk} p^i p^l + \delta^{jl} p^i p^k).  \label{Eq:tensor-I-2}
\end{eqnarray}
The coefficients of each term on the left hand side of the above equations can be determined by constructing the corresponding projection operators:
\begin{eqnarray}
    Q_i^{a_0} &=& -\frac{1\,}{|{\bf p}|^2}\, p_i \nonumber \\
    Q _{ij}^{b_0} &=& \frac{1}{2 |{\bf p}|^2} \left [-|{\bf p}|^2 \delta_{ij} + p_i p_j \right]  \nonumber\\
    Q _{ij}^{b_1} &=& \frac{1}{2 |{\bf p}|^4} \left [-|{\bf p}|^2 \delta_{ij} + 3p_i p_j \right] \nonumber \\
    Q_{ijk}^{c_0} &=& \frac{1}{6|{\bf p}|^4} \left [ |{\bf p}|^2 (\delta_{ij} p_k + \delta_{ik} p_j + \delta_{kj} p_i) - 3 p_i p_j p_k \right]  \nonumber \\
    Q_{ijk}^{c_1} &=& \frac{1}{2|{\bf p}|^6} \left [ |{\bf p}|^2 (\delta_{ij} p_k + \delta_{ik} p_j + \delta_{kj} p_i) - 5 p_i p_j p_k \right]  \nonumber \\
    Q_{ijkl}^{d_0} &=& \frac{1}{24 |{\bf p}|^4} \Big[|{\bf p}|^4 (\delta_{ij} \delta_{kl} + \delta_{ik} \delta_{jl} + \delta_{il}\delta_{jk}) + 3p_i p_j p_k p_l \nonumber\\
 &&-|{\bf p}|^2(\delta_{ij} p_k p_l + \delta_{kl} p_i p_j + \delta_{ik} p_j p_l  \nonumber \\
    &&  + \delta_{il} p_j p_k + \delta_{jk} p_i p_l + \delta_{jl} p_i p_k)\Big] \nonumber \\
    Q_{ijkl}^{d_1} &=& \frac{1}{8 |{\bf p}|^8} \Big[|{\bf p}|^4 (\delta_{ij} \delta_{kl} + \delta_{ik} \delta_{jl} + \delta_{il}\delta_{jk}) + 35 p_i p_j p_k p_l \nonumber\\
 &&-5|{\bf p}|^2(\delta_{ij} p_k p_l + \delta_{kl} p_i p_j + \delta_{ik} p_j p_l  \nonumber \\
    &&  + \delta_{il} p_j p_k + \delta_{jk} p_i p_l + \delta_{jl} p_i p_k)\Big] \nonumber \\
     Q_{ijkl}^{d_2} &=& \frac{1}{24 |{\bf p}|^6} \Big[|{\bf p}|^4 (\delta_{ij} \delta_{kl} + \delta_{ik} \delta_{jl} + \delta_{il}\delta_{jk}) + 15 p_i p_j p_k p_l \nonumber\\
 &&-3|{\bf p}|^2(\delta_{ij} p_k p_l + \delta_{kl} p_i p_j + \delta_{ik} p_j p_l  \nonumber \\
    &&  + \delta_{il} p_j p_k + \delta_{jk} p_i p_l + \delta_{jl} p_i p_k)\Big] \label{Eq:projection} 
\end{eqnarray}
Now, to determine the coefficients, we contract each projection operator with the tensor of same rank from Eqs.~\eqref{Eq:tensor-I-2}. 
Using Eqs.~\eqref{Eq:tensor-I-2} and \eqref{Eq:projection}, the dilepton rate Eq.~\eqref{Eq:dil-decompose} can be rewritten as
\begin{widetext}
\begin{eqnarray}
    \frac{dR^{(2)}}{d^4p} &=& \frac{\beta}{\beta_\pi} \Bigg\lbrace 
    \frac{5}{7 \beta_\pi} \left[\left(Q_{ij}^{b_0} I^{ij} \right) \delta^{ij} + \left(Q_{ij}^{b_1} I^{ij} \right) p^i p^j \right] \pi_i^k \pi_{jk} 
    - \frac{2\tau_\pi}{3} \left(Q_{ij}^{b_1} I^{ij} \right) p^{ij} \pi_{ij} \Theta - \frac{E}{70 \beta_\pi} \pi_{ij} \pi^{ij} \nonumber \\
    && - \frac{2\tau_\pi}{5} \left(Q_{i}^{a_0} I^i \right) p^i \nabla^j \pi_{ij} 
    - {\tau_\pi} \left[\left(Q_{ijk}^{c_0} I^{ijk} \right) (\delta^{ij} p^k + \delta^{ik} p^j + \delta^{kj} p^i) + \left( Q_{ijk}^{c_1} I^{ijk} \right) p^i p^j p^k \right] \nabla_k \pi_{ij} \nonumber \\
    && + \frac{1}{2\beta_\pi} \Big[\left(Q_{ijkl}^{d_0} I^{ijkl} \right) (\delta^{ij} \delta^{kl} + \delta^{ik} \delta^{jl} + \delta^{il}\delta^{jk}) + \left(Q_{ijkl}^{d_1} I^{ijkl} \right) p^i p^j p^k p^l + \left(Q_{ijkl}^{d_2} I^{ijkl} \right) (\delta^{ij} p^k p^l \nonumber \\
    &&  + \delta^{kl} p^i p^j + \delta^{ik} p^j p^l + \delta^{il} p^j p^l + \delta^{jk} p^i p^l + \delta^{jl} p^i p^k)  \Big]\pi_{ij} \pi_{kl}
    \Bigg\rbrace.  \label{Eq:dil-decompose-2}
\end{eqnarray}
\end{widetext}
Substituting Eq.~\eqref{Eq:tensor-I} in Eq.~\eqref{Eq:dil-decompose-2} and following the calculations of Refs.~\cite{Vogt:2007zz,Bhatt:2011kx}, we derive the rate of dilepton production up to second-order CE correction. In a general frame with four-velocity $u^\mu$, the rate expression can be now obtained as
\begin{widetext}
\begin{eqnarray}
 \frac{dR^{(2)}}{d^4p} &=& \frac{dR^{id}}{d^4p} \Bigg\{ 
 \frac{5\beta}{7\beta_\pi^2} \Bigg[
 \frac{1}{2 \mathscr{E}^3} \left(-\frac{\mathscr{E} (u \cdot p)}{2}M^2 + \frac{M^4}{4} \mathscr{Y} \right) \pi_{\beta \gamma} \pi^{\beta \gamma} + \frac{1}{2\mathscr{E}^5} \left( \frac{\mathscr{E} (u \cdot p)}{2} (2 \mathscr{E}^2 - 3M^2) + \frac{3}{4}M^4 \mathscr{Y} \right) p^\alpha p^\beta
 \pi_\alpha^\gamma \pi_{\beta \gamma} \Bigg ]  \nonumber \\
 && - \frac{\beta\tau_\pi}{3\beta_\pi} \frac{1}{\mathscr{E}^5}\left( \frac{\mathscr{E} (u \cdot p)}{2} (2 (u \cdot p)^2-5M^2) + \frac{3}{4}M^4 \mathscr{Y} \right)
 p^\alpha p^\beta \pi_{\alpha\beta}\Theta - \frac{\beta}{70\beta_\pi^2} (u \cdot p) \pi^{\alpha\beta} \pi_{\alpha\beta} 
 - \frac{\beta\tau_\pi}{5\beta_\pi}p^\alpha \nabla^\beta \pi_{\alpha\beta} \nonumber \\
 && - \frac{\beta \tau_\pi}{\beta_\pi} \Bigg[ \frac{1}{2 \mathscr{E}^5} \left(-M^2 \mathscr{E} ((u \cdot p)^2 + 2M^2) + \frac{3}{2} (u \cdot p) M^4 \mathscr{Y} \right) g^{\alpha\gamma} p^\beta  \nonumber \\
 &&+ \frac{1}{2 \mathscr{E}^7} \left((5 (u \cdot p)^2 - 4 \mathscr{E}^2) \mathscr{E}^3 - \frac{15}{2} (u \cdot p)^2 M^2 \mathscr{E} + \frac{15}{4} (u \cdot p) M^4 \mathscr{Y} \right) p^\alpha p^\beta p^\gamma 
 \Bigg] \nabla_\gamma \pi_{\alpha\beta} \nonumber \\
 && + \frac{\beta}{16\beta_\pi^2} \frac{1}{ \mathscr{E}^5 } \Bigg[2 M^4 \Bigg(\beta\Big[ \frac{\mathscr{E}}{3} ((u \cdot p)^2 + 2M^2) - \frac{M^2 (u \cdot p)}{2} \mathscr{Y} \Big]
 +\Big[-3 \mathscr{E} (u \cdot p) + \frac{1}{2} (2 (u \cdot p)^2 + M^2) \mathscr{Y} \Big]\Bigg) g^{\alpha\mu}g^{\beta\nu}\nonumber \\
 &&+\frac{1}{\mathscr{E}^4} \Bigg(\beta\Bigg[\mathscr{E} (u \cdot p)^2 \Big(\frac{8}{3} (u \cdot p)^4 + 29M^4 - \frac{38}{3} (u \cdot p)^2 M^2 \Big) + 16 \mathscr{E} M^6 - \frac{35}{2} (u \cdot p) M^6 \mathscr{Y} \Bigg]\nonumber\\
 &&+\Bigg[\mathscr{E} (u \cdot p) (4 (u \cdot p)^4 - 81M^4 - 28 (u \cdot p)^2 M^2) + \frac{15}{2}M^4(6 (u \cdot p)^2 + M^2) \mathscr{Y} \Bigg]\Bigg) p^\alpha p^\beta p^\mu p^\nu\nonumber\\
 &&+\frac{4M^2}{\mathscr{E}^2}\Bigg(\beta \Bigg[\mathscr{E}\Big(-\frac{2}{3} (u \cdot p)^4 + \frac{8}{3}M^4 + 3 (u \cdot p)^2 M^2 \Big) - \frac{5}{2} (u \cdot p) M^2 \mathscr{Y} \Bigg]\nonumber\\
 &&+ \Bigg[-\mathscr{E} (u \cdot p) (2 (u \cdot p)^2 + 13 M^2) + \frac{3}{2} M^2 (M^2 + 4(u \cdot p)^2) \mathscr{Y} \Bigg] \Bigg) g^{\alpha\mu}p^\beta p^\nu \Bigg] \pi_{\alpha\beta}\pi_{\mu\nu}
 \Bigg\}.  \label{Eq:dil-rate-IIorder}
\end{eqnarray}
\end{widetext}
The total dilepton rate in presence of second-order correction (Eq.~\eqref{Eq:total-rate-expression}) is given by sum of the ideal~\eqref{Eq:dil-rate-ideal}, first-order~\eqref{Eq:dil-rate-Iorder} and second-order~\eqref{Eq:dil-rate-IIorder} contributions. 

Further, one can obtain the shear viscous contribution to the dilepton rate due to Grad's ansatz as~\cite{Dusling:2008xj}
\begin{eqnarray}
    \frac{dR^G}{d^4p} = \frac{dR^{id}}{d^4p}\frac{2}{3}\left(\frac{p^\mu p^\nu}{2sT^3}\pi_{\mu\nu}\right),  \label{Eq:dil-grad}
\end{eqnarray}
so that the total rate becomes  $dR/d^4p = dR^{id}/d^4p + dR^G/d^4p$. 

\subsection{Phase-space weighted rates}

\begin{figure*} 
   \centering
     \subfigure
     []{
    \includegraphics[width=0.49\textwidth]{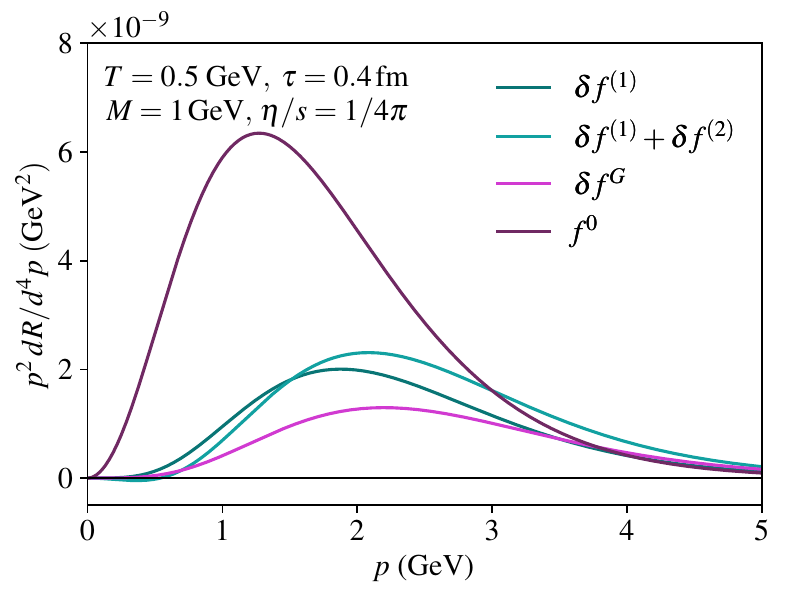}\label{fig:dil-rate-1}} 
    \subfigure[]{\includegraphics[width=0.49\textwidth,height=6.7cm]{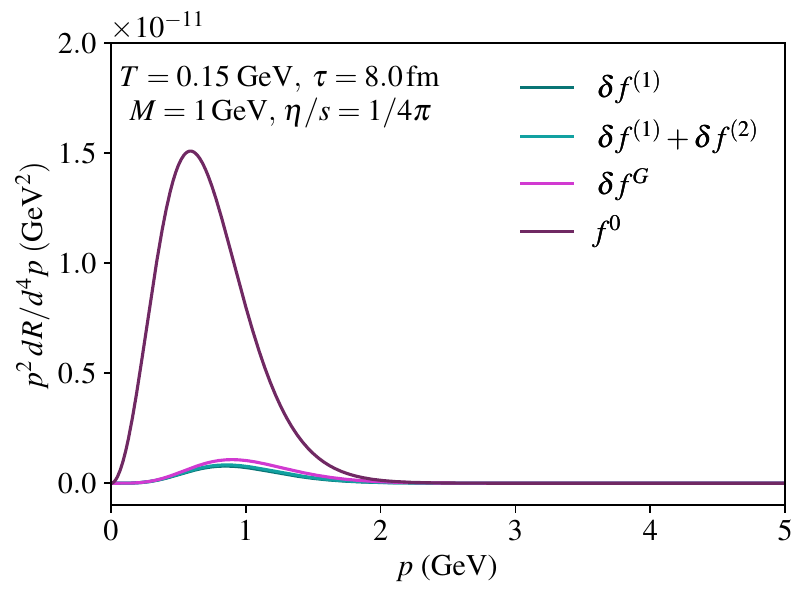}\label{fig:dil-rate-1}}
  \caption{Phase-space weighted thermal dilepton rates considering contributions from first-order $\delta f^{(1)}$ and up to second-order $\delta f^{(1)} + \delta f^{(2)}$ Chapman-Enskog, and Grad's $\delta f^G$ viscous corrections, for early and late time conditions, within Bj\"orken flow. We fix $M=1$ GeV, $\eta/s = 1/4\pi$. The ideal contribution is also shown.}
  \label{fig:dil-rate-phase}
  \end{figure*}

In order to analyse the impact of viscous corrections on the dilepton rate, we plot the phase-space weighted rate $p^2 dR/d^4 p$ as a function of dilepton momentum $p$ in Fig.~\ref{fig:dil-rate-phase}, considering individual viscous contributions in the distribution functions, within Bj\"orken flow. In the rate expressions obtained earlier, we use the viscous corrections given in Sec.~\ref{sec:momentum-dep-deltaf} and consider $p_z=0$ with the Navier-Stokes value for shear stress. We take $M=1$ GeV, $\eta/s = 1/4\pi$, and do this analysis for representative early and late-time conditions of the evolution. 
We observe that the contribution to the rate arising from the $\delta f^{(1)}$ term increases from zero and has a peak in the intermediate momentum range, then decreases with $p$. When second order terms of CE $\delta f^{(2)}$ are incorporated, the corrections decrease in the region $p<2$ GeV, while they become larger than $\delta f^{(1)}$ contribution for $p> 2$ GeV. 
The contribution of Grad's $\delta f^G$ to the rate has a broad momentum dependence and remains small compared to that of CE corrections up to $p=4$ GeV; thereafter, it becomes larger than the $\delta f^{(1)}$ modification. From the late time snapshot given, we observe that the contribution from $\delta f^G$ becomes dominant owing to its quadratic momentum structure. 
This shift in the behaviour of $\delta f^G$ may have effect on the final dilepton spectra. We also plot the rates neglecting the corrections (considering the ideal distribution) for both the system conditions, and it is seen that at high temperature, the dilepton rates from all the corrections are larger at high momentum compared to the ideal scenario, indicating the violation of $\delta f_i/f^0 <<1$ in that region. All these features remain qualitatively same at late times, though the magnitude of the corrections decreases. 
\subsection{Dilepton yields}

We determine the thermal dilepton spectra from heavy-ion collisions by convoluting the total dilepton rate over the space-time history of the collisions. The components of particle four-momentum $p^\mu$ in Milne coordinates are given by 
$p^\mu = (m_T \cosh(y-\eta_s), p_T \cos\phi_p, p_T\sin\phi_p, m_T \sinh(y-\eta_s))$; where $\phi_p$ is the azimuthal angle and $y$ is the particle rapidity. Here, $m_T^2 = p_T^2 + M^2$, with $p_T$ and $M$ being the transverse momentum and invariant mass of the particle. Noting the four-velocity of the fluid to be $u^\mu = (1, 0, 0, 0)$, we obtain the terms appearing in the viscous corrections in the Milne coordinates within Bj\"orken flow as
\begin{eqnarray}
\pi^{\alpha\beta}\pi_{\alpha\beta}&=&\frac{3}{2} \pi^2 \nonumber\\
p^{\alpha}p^{\beta}\pi_{\alpha\beta}& = & \left[\frac{p^2_{T}}{2} - m_T^2 \sinh^2(y-\eta_{s}) \right]\pi \nonumber\\
p^{\alpha}p^{\beta}\pi^{\gamma}_{\alpha}\pi_{\gamma\beta}&=&
- \left[ \frac{p^2_{T}}{4} + m^2_{T} \sinh^2(y-\eta_{s}) \right] \pi^2  \nonumber\\
p^{\alpha}\nabla^{\beta}\pi_{\alpha\beta}&=&- m_T \cosh(y-\eta_{s}) \frac{\pi}{\tau} \nonumber\\
p^{\alpha}p^{\beta}p^{\gamma}\nabla_{\alpha}\pi_{\beta\gamma}&=& 
m_T^3 \sinh(y-\eta_{s})\sinh2(y-\eta_{s}) \frac{\pi}{\tau}. \nonumber
\label{Eq:terms-milne}
\end{eqnarray}
Noting the four-volume element within Bj\"orken flow as $d^4 x = \pi R_A^2 \tau d\tau d\eta_s$, where $R_A = 1.2 A^{1/3}$ fm is the radius of colliding nuclei, we write the dilepton yield from QGP as

\begin{align}
\frac{d N}{ d M^2 d^2 p_{T} d y} &= 
\pi R_{A}^{2} 
\int_{\tau_{0}}^{\tau_f} d \tau \tau \int_{-\infty}^{\infty} d \eta_{s}
\left(\frac{1}{2} \frac{d R}{d^{4} p}\right). 
\end{align}
Here, $\tau_f$ is the proper-time at which the temperature of the system approaches the critical temperature $T_c$.
The thermal dilepton yield from non-viscous QGP can be further simplified, by employing the ideal Bj\"orken solution $T(\tau) = (\tau_0/\tau)^{1/3} T_0$, as 
 \begin{equation}\label{dilid-yield}
 \frac{d N^{id}}{d M^2 d^2 p_{T} d y} = 2\mathscr{F} 
 \int_{\tau_0}^{\tau_f}d\tau\,\tau K_0(z_T); 
\end{equation}
where, $z_T \equiv m_T/T$ and $\mathscr{F} = 5R_A^2 \alpha_e^2/(72 \pi^3) $. 
\begin{eqnarray}
\frac{d N^{id}}{d M^2 d^2 p_{T} d y} &=& 6 \mathscr{F} \frac{\tau_0^2}{z_0^6} \Bigg[ z_T \Big[ (4 z_T^3 + 32 z_T) K_0(z_T) \nonumber \\
&&+ (z_T^4 + 16 z_T^2 +64) K_1(z_T)\Big]\Bigg]_{z_f}^{z_0};
\end{eqnarray}
where $z_0 = m_T/T_0$ and $z_f = m_T/T_c$.

\section{Results and Discussion}
\label{sec:results}

\begin{figure*} 
   \centering
     \subfigure
     []{
    \includegraphics[width=0.49\textwidth]{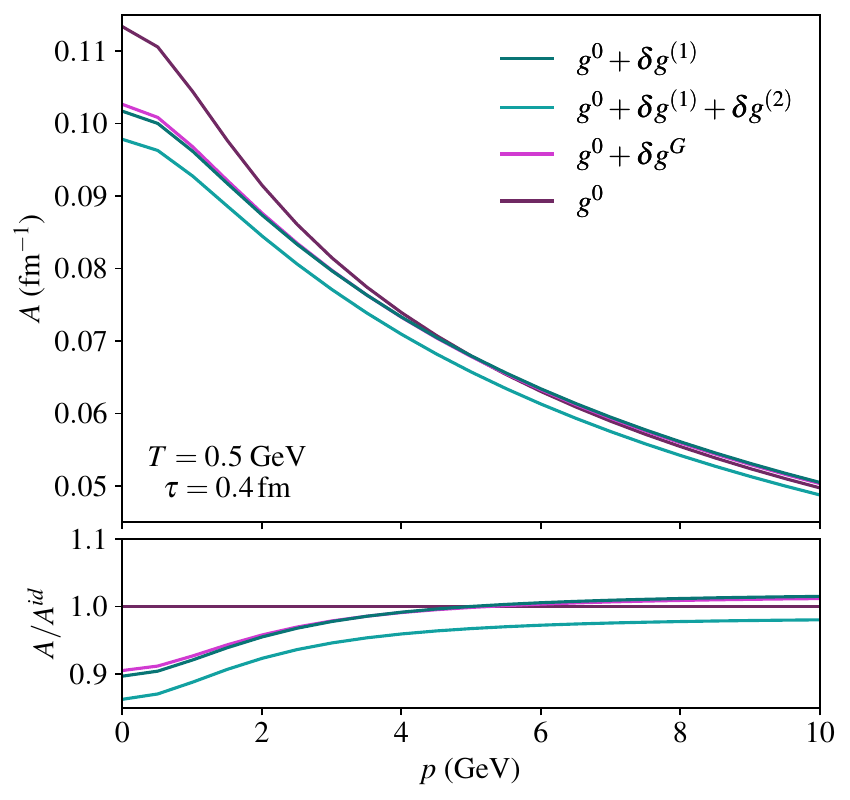}\label{fig:A-p-initial}} 
    \subfigure[]{\includegraphics[width=0.49\textwidth,height=8.3cm]{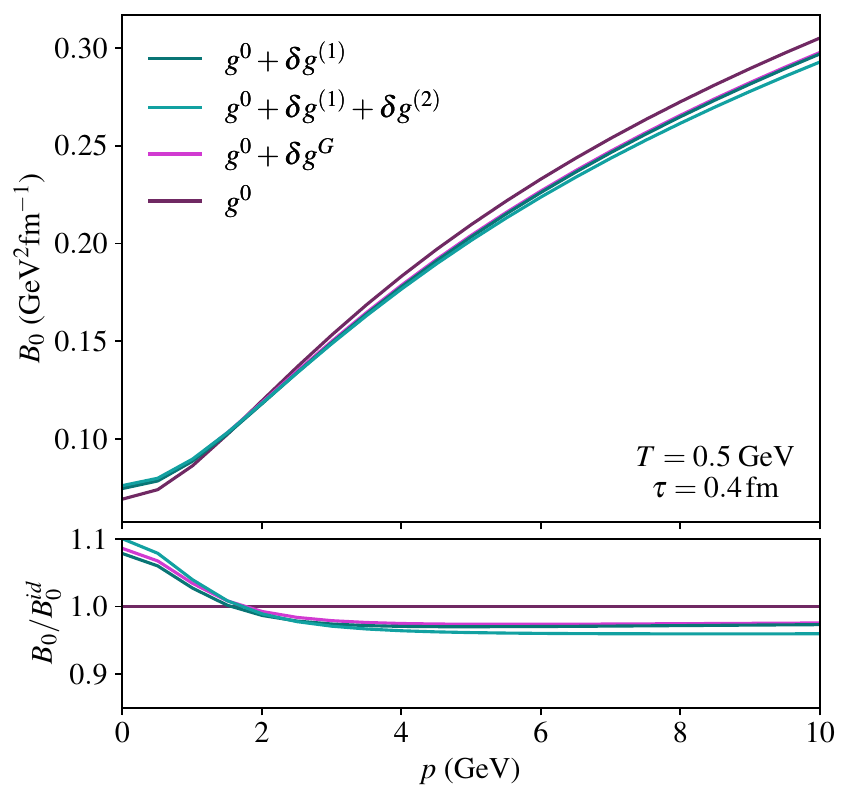}\label{fig:B0-p-initial}}
  \caption{(a) Drag ($A$) and (b) transverse diffusion ($B_0$) coefficients of heavy quark as a function of momentum $p$ in presence of different viscous corrections, for early time of evolution. The coefficients in the absence of viscosity are also shown. The lower panels depict the ratio of the coefficients in presence of viscosity to that determined in the ideal case. 
  }
  \label{fig:A-B0-p}
  \end{figure*}

\begin{figure}
    \centering
    \includegraphics[width=0.49\textwidth]{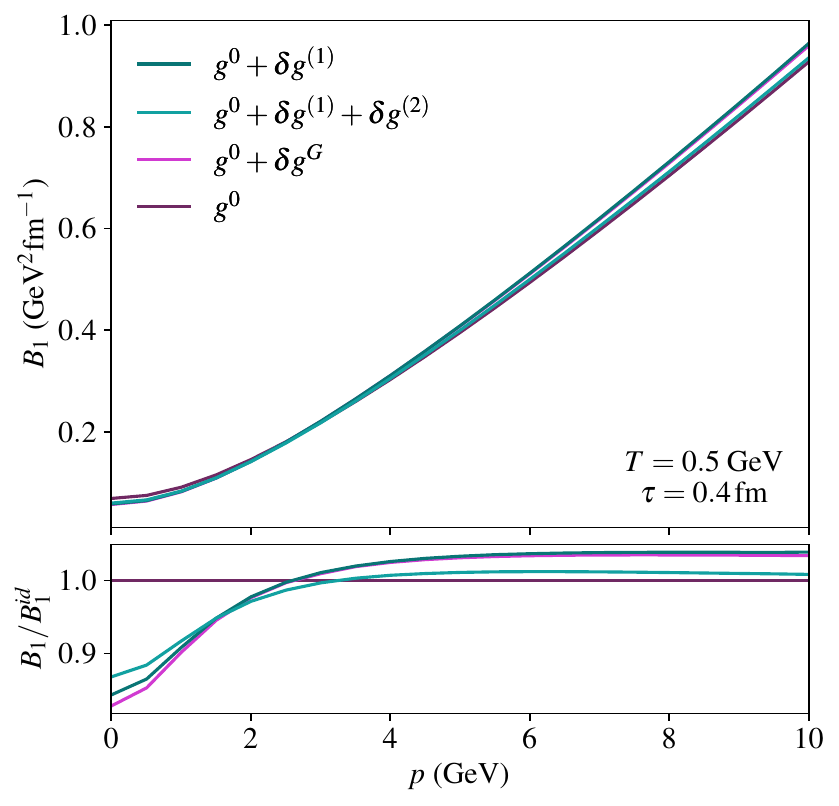}
    \caption{Longitudinal diffusion coefficient ($B_1$) of heavy quark as a function of momentum $p$ in presence of different viscous corrections, for early time of evolution. The coefficient in the absence of viscosity is also shown. The lower panels depict the ratio of the coefficient in presence of viscosity to that obtained in the ideal case. }
    \label{fig:B1-p}
\end{figure}

\begin{figure*} 
   \centering
     \subfigure
     []{
    \includegraphics[width=0.49\textwidth, height=8.7cm]{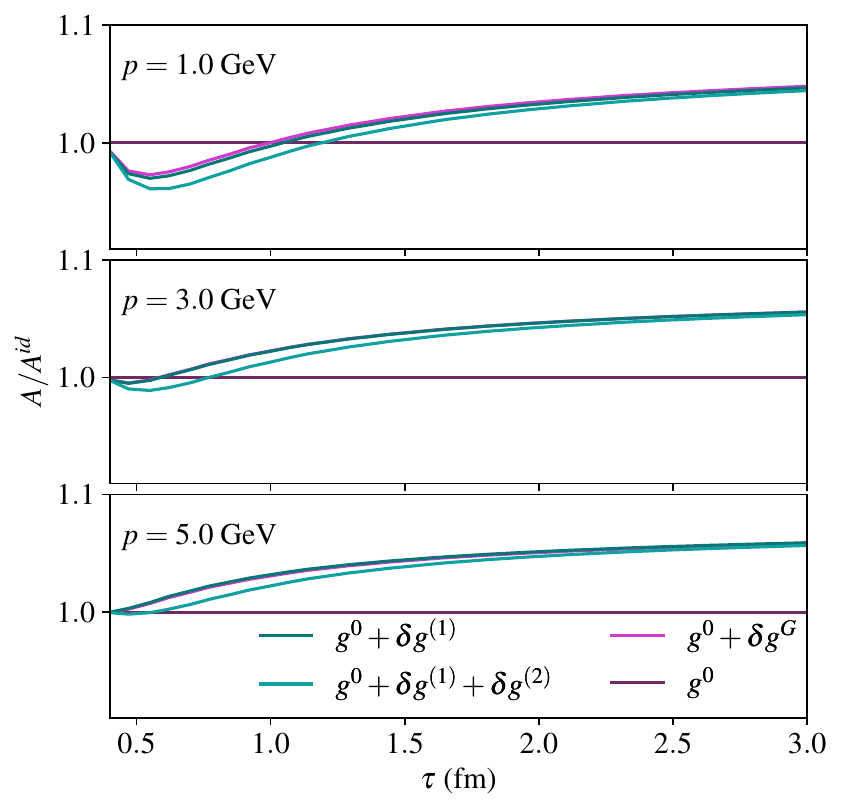}\label{fig:A-tau}} 
    \subfigure[]{\includegraphics[width=0.49\textwidth, height=8.7cm]{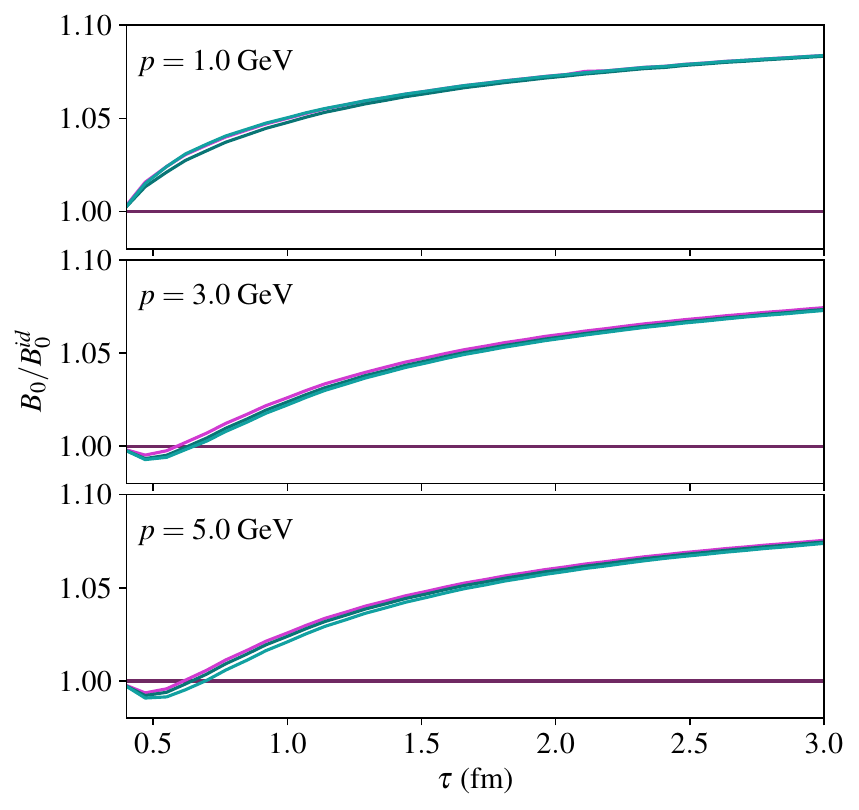}\label{fig:B0-tau}}
  \caption{Temporal evolution of (a) drag and (b) transverse diffusion coefficients of heavy quark varying its momentum $p$ using different forms of viscous correction.}
  \label{fig:A-B0-tau}
  \end{figure*}

\begin{figure}
    \centering
    \includegraphics[width=0.49\textwidth]{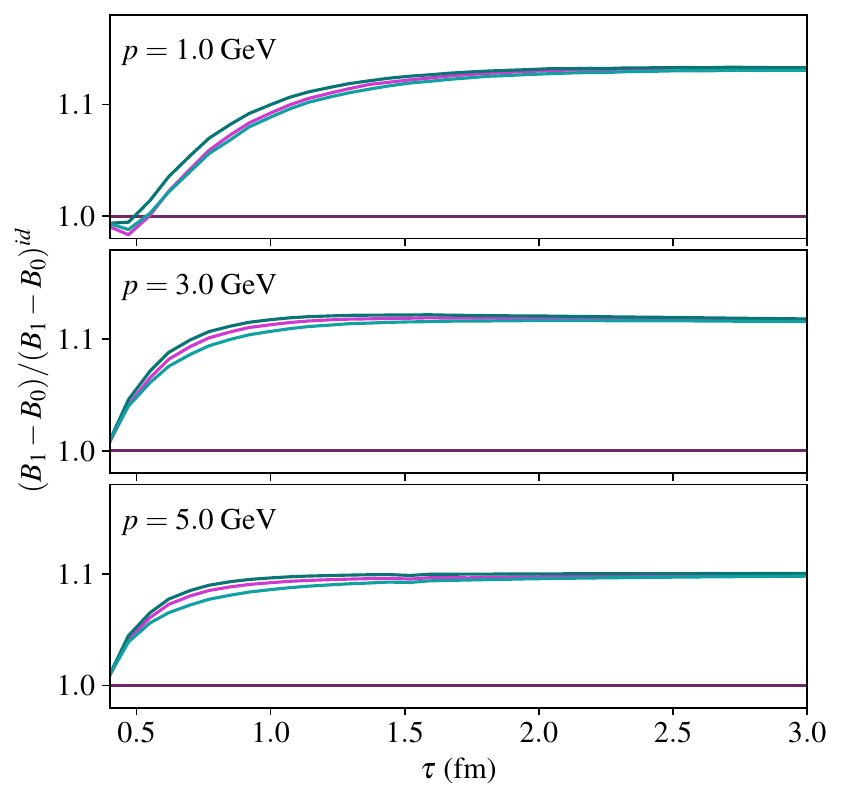}
    \caption{Measure of anisotropy in momentum diffusion using different viscous corrections by varying HQ momentum as a function of proper time. The legends corresponding to different curves are the same as those used in Fig.~\ref{fig:A-tau}.
    }
    \label{fig:B10-tau}
\end{figure}

\begin{figure*} 
   \centering
     \subfigure
     []{
    \includegraphics[width=0.49\linewidth]{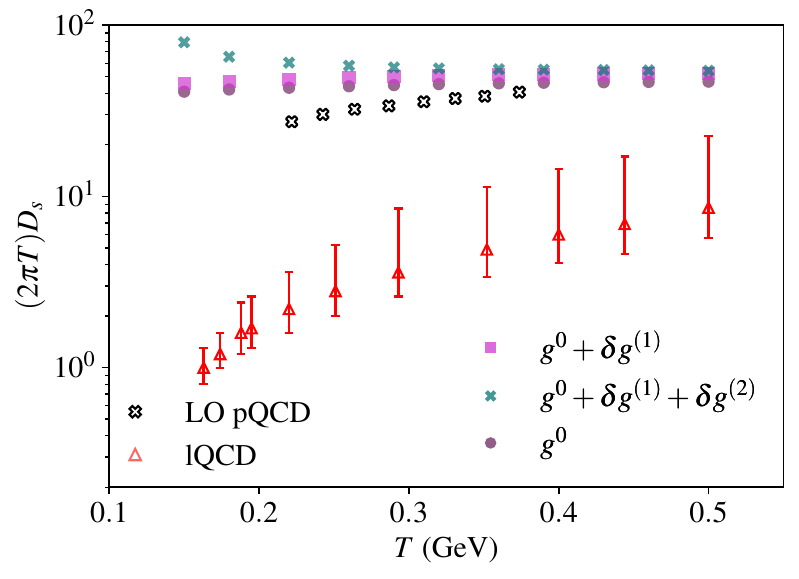}\label{fig:Ds-T}} 
    \subfigure[]{\includegraphics[width=0.49\linewidth]{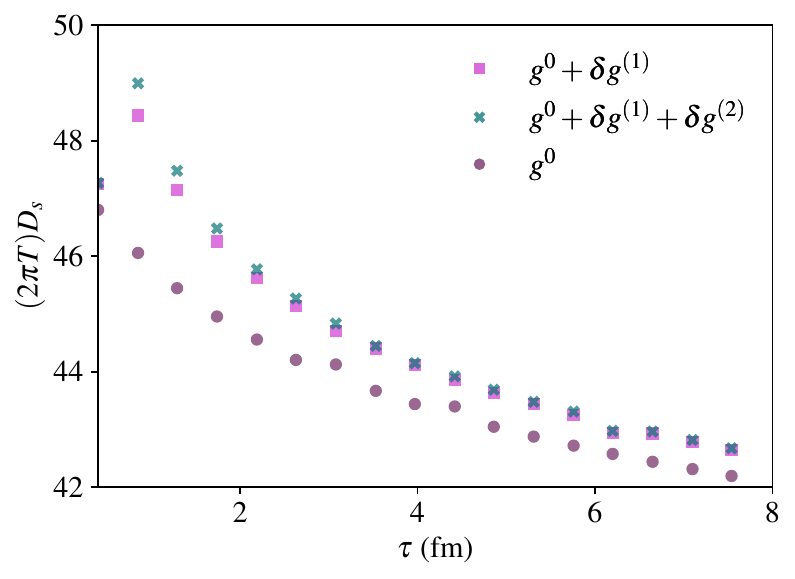}\label{fig:Ds-tau}}
  \caption{Scaled spatial diffusion coefficient as a function of (a) temperature  and (b) proper time, for CE corrections up to first- and second-orders. The results from leading order perturbative QCD (LO pQCD)~\cite{vanHees:2004gq} and lattice QCD (lQCD)~\cite{HotQCD:2025fbd} are also depicted. }
  \label{fig:Ds}
  \end{figure*}

In this section, we study the heavy-quark transport and thermal dilepton spectra with different forms of correction from expanding QGP by numerically solving the relevant equations. 
Unlike previous studies restricted to first order viscous corrections, the present analysis includes the complete second order Chapman-Enskog contribution for both heavy quark transport and thermal particle  production. 
We employ the temperature and shear stress evolution profiles obtained by numerically solving the respective dynamical equations, Eqs.~\eqref{Eq:epsilon-evolution} and \eqref{Eq:pi-evolution}, discussed in Sec.~\ref{sec:bjorken}. We take the initial conditions for Au-Au ($A=197$) collisions at RHIC: $T(\tau_0) \equiv T_0 = 0.5$ GeV, $\pi(\tau_0)\equiv \pi_0=0$ at $\tau_0 = 0.4$ fm, and we fix the value of shear viscosity as $\eta/s = 1/4\pi$. The critical temperature is chosen to be $T_c = 0.15$ GeV, and the corresponding proper-time for the viscous evolution is determined as $\tau_f=16.7$ fm. For the $\delta f_i =0$ case, we use the ideal Bj\"orken temperature profile $T (\tau) = (\tau_0/\tau)^{1/3} T_0$, for which we get $\tau_f=14.8$ fm. Also, the dilepton spectra is plotted for midrapidity region, $y=0$. 
\par
First we look into the heavy quark drag ($A$) and diffusion coefficients ($B_0$ and $B_1$) in presence of viscous corrections, at an early stage of evolution ($T=0.5$ GeV, $\tau=0.4$ fm) in Figs.~\ref{fig:A-B0-p} and \ref{fig:B1-p}. The upper panels display momentum dependence of drag and diffusion coefficients, while the corresponding lower panels show the ratios with respect to the ideal results. Since the viscous corrections are strongly suppressed at later stages of expansion, only the early-stage results are presented here. 
\par
We study the drag coefficient as a function of HQ momentum $p$ in Fig.~\ref{fig:A-p-initial}. 
We observe that the viscous corrections suppress the drag force in the momentum range $p<4$ GeV with the largest suppression shown by total CE correction $\delta g^{CE}=\delta g^{(1)} + \delta g^{(2)}$, followed by $\delta g^{(1)}$ and $\delta g^G$. When $p$ is increased, both $\delta g^{(1)}$ and $\delta g^G$ contributions gradually decrease to zero around $p \sim 5$ GeV, and they start to moderately enhance the drag force at higher momenta, with the $\delta g^{(1)}$ correction being slightly large compared to $\delta g^G$. These results are in line with previous studies~\cite{Das:2012ck,Shaikh:2021lka}. Whereas, though the suppression from the total CE contribution reduces with increase in $p$, it still remains as the dominant correction for the entire $p$ range. This behaviour of the drag coefficient can be understood from the corresponding kernel analysis (Fig.~\ref{fig:A-kernel}). The dominant contribution to the drag coefficient originates from thermal partons with momenta $q \sim (1-2)$ GeV, where the equilibrium kernel attains its maximum. The CE correction exhibits the largest modification precisely in this momentum region, resulting in a stronger overlap with the drag kernel. Consequently, the integrated viscous correction to the drag coefficient is significantly larger in the second-order CE framework and results in suppression throughout the momentum range. 
\par
Now, we show the transverse diffusion coefficient $B_0$ as a function of HQ momenta, using different viscous corrections in Fig.~\ref{fig:B0-p-initial}. For small HQ momenta, the fluctuations in momentum transfer increase in presence of viscosity, and they decrease with increment in $p$ and around $p\sim 2$ GeV, start to give moderate suppression. Throughout the momentum range, 
viscous modifications are larger for $\delta g^{CE}$ followed by $\delta g^{G}$ and $\delta g^{(1)}$, respectively. 
Coming to the longitudinal diffusion coefficient $B_1$ in Fig.~\ref{fig:B1-p}, 
the impact of viscous terms is to suppress $B_1$ at small $p$, and to enhance for $p>3$ GeV. We find that the contribution from total CE correction $\delta g^{CE}$ remains subleading at almost all momenta compared to $\delta g^{G}$, unlike the previous coefficients. These results demonstrate that the final modification of a transport coefficient is controlled not only by the magnitude of the non-equilibrium correction but also by its momentum-space overlap with the underlying transport kernel.
\par
Next, we look into the temporal evolution of transport coefficients of HQ at different momenta in Figs.~\ref{fig:A-B0-tau} and \ref{fig:B10-tau}, by plotting the ratio of the coefficients with respect to the ideal contributions. Note that we have included the temperature and shear stress evolution profiles in this analysis. The transport coefficients decrease monotonically with $\tau$ due to the cooling of the system. From Fig.~\ref{fig:A-tau}, for small HQ momenta, it is observed that the viscous corrections suppress the drag at early times, and this suppression is observed to be higher for $\delta g^{CE}$. With increment in $\tau$, the suppression of drag force weakens, leading to a monotonic enhancement throughout the evolution, with the largest enhancement arising due to $\delta g^G$ correction. For higher values of $p$, the early-time suppression due to viscosity reduces and the total CE correction becomes lesser compared to that of Grad's over the complete evolution. Similarly, we plot the ratio of diffusion coefficient $B_0/B_0^{id}$ as a function of $\tau$ in Fig.~\ref{fig:B0-tau}. In contrast to the previous case, for low $p$, all the corrections enhance $B_0$, while the early-time suppression due to viscosity becomes prominent only at large $p$. However, for all $p$, the largest correction is displayed by $\delta g^{CE}$ at very early times. Next, in Fig.~\ref{fig:B10-tau}, we show the proper-time dependence of the quantity $B_1 - B_0$. For small $p$, the largest suppression is shown by $\delta g^G$ at very early times, and as the system evolves, the corrections begin to enhance $B_1 - B_0$, with contribution from $\delta g^{CE}$ being subleading. We find that with increment in HQ momenta, the corrections enhance $B_1 - B_0$ with lesser modifications from $\delta g^{CE}$ correction. We illustrate that, a distinct separation between the contributions of different corrections can be seen at early time for all the coefficients, and as the system evolves, the relative importance of viscous effects gradually decreases. This feature is consistent with the analysis of viscous corrections and transport kernels presented earlier, which showed that the notable viscous contributions originate from the early, high-temperature stages of the evolution. 
\par
The spatial diffusion coefficient $D_s$ of the heavy quark is plotted in Fig.~\ref{fig:Ds}. We show the temperature dependence of scaled spatial diffusion coefficient in Fig.~\ref{fig:Ds-T}, for $\tau = \tau_0=0.4$ fm, using the CE correction up to first and second orders. The coefficient corresponding to the ideal case is also shown for comparison. It can be seen that the presence of viscous corrections enhances $D_s$ over the entire temperature range. This is in line with the earlier observation of the behaviour of the drag coefficient $A$ in the presence of viscosity (Fig.~\ref{fig:A-p-initial}). The increment is observed to be large for $\delta g^{CE}$, especially near the transition temperature $T_c$. We note that our values of $D_s$ agree with the leading-order pQCD results of Ref.~\cite{vanHees:2004gq}. We also depict the non-perturbative effects to the $D_s$ from lattice QCD~\cite{HotQCD:2025fbd}. In Fig.~\ref{fig:Ds-tau}, we depict the temporal evolution of the quantity $(2\pi T)D_s$. It can be seen that $D_s$ decreases as the system evolves. The enhancement due to viscous corrections is observed to be more significant during the initial stage. Also, we see that the second-order terms increase $D_s$ more prominently during the early time, compared to $\delta g^{(1)}$. We note that this analysis is carried out to investigate the effect of second-order terms on $D_s$. 

\begin{figure*}
    \centering
    \includegraphics[width=\textwidth]{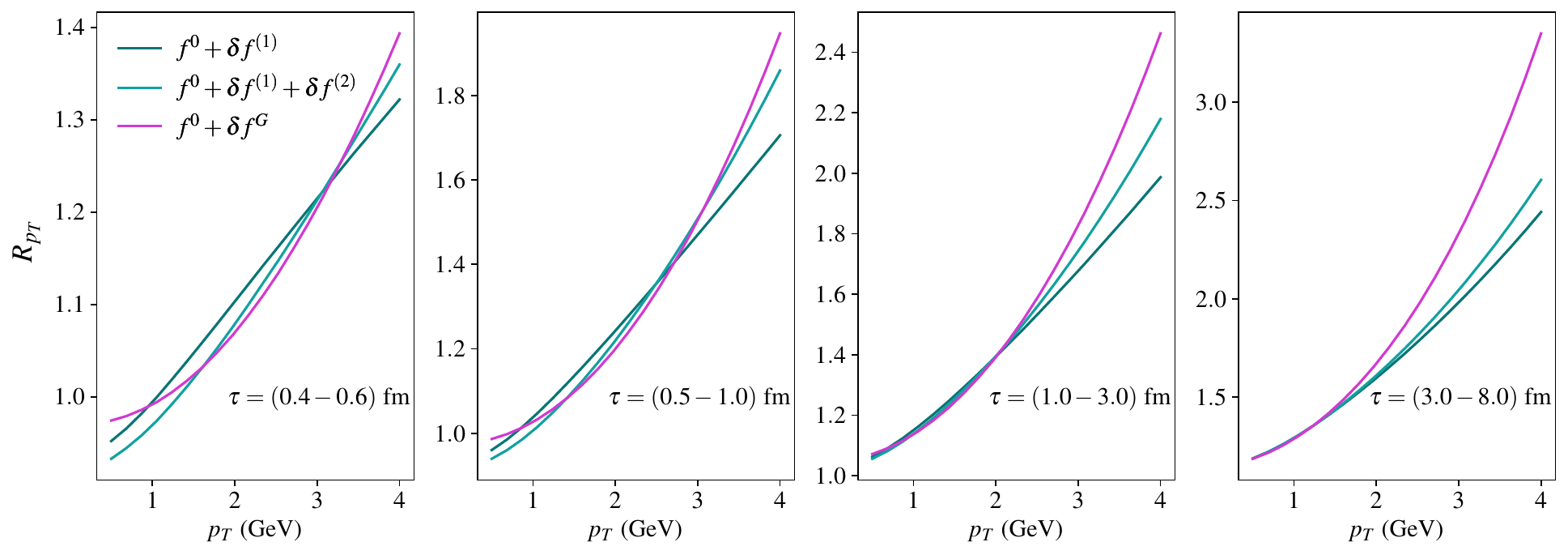}
    \caption{Thermal dilepton spectra (scaled with the ideal contribution) from an expanding QGP for different proper time intervals, using various viscous corrections. The invariant mass of dilepton is fixed to be $M=1$ GeV. }
    \label{fig:dil-spectra-1}
\end{figure*}

\begin{figure}
    \centering
    \includegraphics[width=0.49\textwidth]{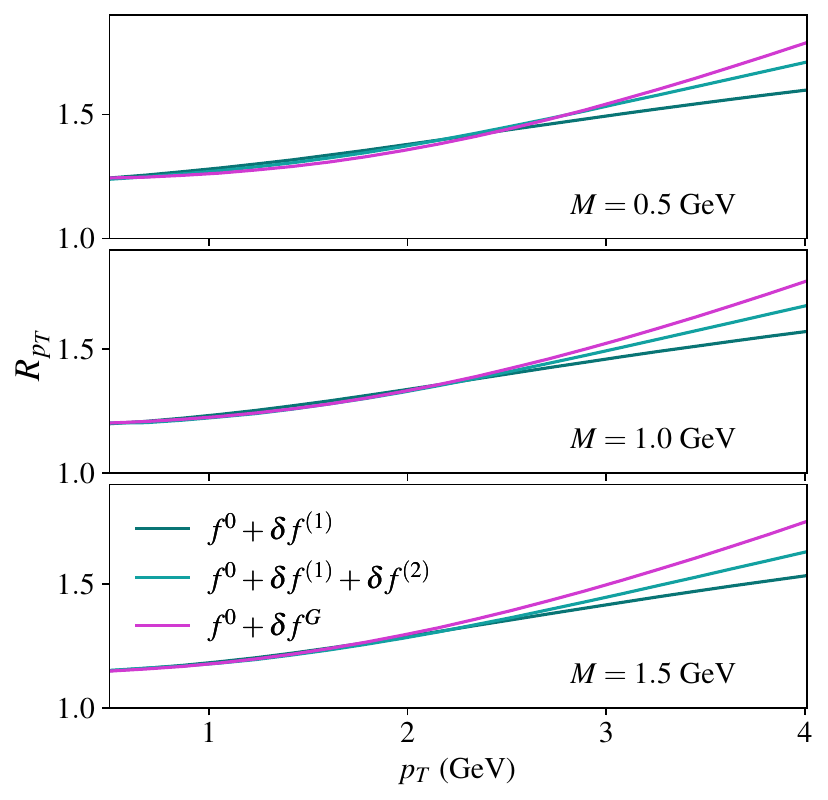}
    \caption{Thermal dilepton spectra from expanding viscous QGP by varying the invariant mass. }
    \label{fig:dil-spectra-2}
\end{figure}

\par
Now, we look into the thermal dilepton spectra using the different forms of viscous corrections. Although viscous corrections are same in both cases, heavy-quark transport probes the instantaneous properties of the medium, whereas dilepton production accumulates contributions over the entire evolution history. Consequently, the observable dependence of viscous effects can differ substantially.
\par
We study the effect of viscous corrections on the dilepton spectra from the expanding QGP in Fig.~\ref{fig:dil-spectra-1}, by plotting the ratio
\begin{eqnarray}
    R_{p_T} = \left[\frac{dN}{dM^2 d^2p_T dy} \right]/\left[ \frac{dN^{id}}{dM^2 d^2p_T dy} \right].
\end{eqnarray}
The strength and behaviour of viscous corrections are analysed by evaluating the dilepton spectra from different proper-time intervals for $M =1$ GeV. In the earliest proper-time bin ($\tau: 0.4-0.6$ fm), the contribution from $\delta f^{(1)}$ correction to the yield is high in the momentum range $1-3$ GeV compared to the total CE contribution. At high $p_T$, however, the cumulative contribution from $\delta f^{CE}$ exceeds that of the first order CE correction $\delta f^{(1)}$. The spectra obtained in the presence of $\delta f^G$ remains subleading in the intermediate $p_T$, while it becomes enhanced compared to that of CE contributions in the ranges $p_T>3$ GeV and $p_T < 1$ GeV. We find that with increment in proper-time, the total CE contribution becomes higher than the first-order correction throughout the $p_T$ range. 
Although this is the scenario, the spectra from Grad's $\delta f^G$ dominates the CE contributions in the momentum region $p_T > 1.5$ GeV. This feature is as seen from the momentum dependence of dilepton rate in Fig.~\ref{fig:dil-rate-phase}, where the Grad's correction increases at later time. As a result, one can expect the dilepton spectra with $\delta f^G$ to be higher at large momenta compared to that obtained using $\delta f^{CE}$, after integrating till $\tau_f$. Also, the strength of $\delta f^{CE}$ lies intermediate to $\delta f^G$ and individual $\delta f^{(1)}$ correction. 

In Fig.~\ref{fig:dil-spectra-2}, we study the total dilepton spectra for different viscous corrections, by varying the dilepton invariant mass $M$. We observe that the dilepton spectra in the presence of total CE correction is less than that obtained with $\delta f^{(1)}$, and around $p_T \sim 2.5$ GeV, $\delta f^{CE}$ overtakes the $\delta f^{(1)}$ contribution to the spectra. For large $M$ values, strength of $\delta f^{CE}$ on the yield is seen to be higher at lower momentum. 
Further, across all $M$, the spectra in presence of $\delta f^G$ is less compared to $\delta f^{CE}$ contributions, while it dominates the high $p_T$ regime. Also, this shift in $\delta f^G$ contribution occurs at an early $p_T$ for higher invariant mass case. 

We note that, the strong gradients and high temperature of the system enhance the dilepton production in presence of different viscous corrections especially at large $p_T$. As the system evolves, the relative importance of the Chapman-Enskog corrections decreases, while the Grad's correction amplifies the high $p_T$ dilepton production due to its stronger momentum dependence. This shift occurs within a very short proper time itself, since the initial cooling of the high temperature medium happens quickly and the system remains in the intermediate temperature for larger times. 
\par 
A comparison of the heavy quark and thermal dilepton results studied reveals that, though the Chapman-Enskog corrections generally produce larger modifications to the thermal distributions compared to the Grad's ansatz, the corresponding impact on physical observables depends on how these distributions are sampled by the observable under consideration. For HQ transport coefficients, the dominant contribution arises from the intermediate momentum region of the thermal partons, where the CE corrections exhibit the largest enhancement and as a result, the drag and diffusion coefficients display a stronger sensitivity to CE corrections. 
Whereas, dilepton production is 
sensitive to a broader momentum range and contains substantial contributions from the high-momentum tail of the distribution. Hence, the observed hierarchy of distribution functions does not directly translate to the dilepton spectra, leading to a qualitatively different dependence of the observable. Thus the final outcome of viscous effects is determined by the interplay between the momentum dependence of the viscous correction and the kernel associated with the observable considered, rather than by the magnitude of $\delta f$ alone. Further, it is seen that the second-order Chapman-Enskog corrections to thermal particle production remain controlled and well behaved, as seen from the earlier observations for hadronic spectra~\cite{Bhalerao:2013pza}. However, in the HQ sector, the transport kernels can substantially amplify the contribution of higher-order viscous corrections.

\section{Summary}
\label{sec:summary}

In summary, we have presented the first study of heavy quark (HQ) transport coefficients and thermal dilepton production in the QGP incorporating viscous corrections up to second order in gradients, obtained using the Chapman-Enskog (CE) method. Beyond quantifying the magnitude of these corrections, we have demonstrated that their manifestation at the observable level is controlled by the momentum-space overlap between the non-equilibrium distribution and the corresponding observable kernel.
\par
First, we analysed how the first- and second-order CE corrections modify the momentum dependence of the distribution function within the Bj\"orken's prescription. We also compared the results with the momentum structure of the ansatz determined using the Grad's 14-moment method. It is observed that the CE corrections dominate in the intermediate momenta, while the Grad's correction begins to dominate with increment in momentum. 
Further, we have employed the causal relativistic dissipative hydrodynamics within Bj\"orken flow to model the evolution of QGP with $\eta/s = 1/4\pi$ and studied how the viscous effects permeates into the observables, namely heavy quark transport coefficients and thermal dileptons. 
\par
The heavy quark transport in hot QGP has been modelled within the framework of Fokker-Planck dynamics. Using the first- and second-order CE corrections to the constituent particle distribution functions in the collision kernel of $2 \rightarrow 2$ scattering processes in the medium, we calculated the viscous-modified HQ drag and momentum diffusion coefficients. We studied the momentum dependence of HQ transport coefficients and found that the contribution due to the second-order CE correction remains significant. Importantly, we observed that inclusion of CE correction up to the second-order substantially modifies the drag force by suppressing it throughout the momentum range, while the first-order correction enhances the drag at large momenta. We found that viscous effects enhance the transverse diffusion at very low momenta and decrease it otherwise, with maximum modification arising from the total CE correction. On the other hand, the longitudinal diffusion in presence of viscosity appears less than that of the ideal case at low momenta and starts to increase with increment in the momentum, with subleading contribution from total CE correction. 
We also looked into the evolution of transport coefficients of HQ by varying the momentum. We found that the second-order corrections are important and are significant at initial proper times. The spatial diffusion coefficient of HQ was studied and compared with the leading order perturbative QCD and lattice QCD results. This analysis demonstrated that the inclusion of second-order terms in CE correction results in a large enhancement of the coefficient near the transition temperature. 
\par
Next, by using the viscous-modified parton distribution functions, we calculated the thermal dilepton production rate analytically up to the second-order CE correction. The phase-space weighted rates were analyzed to understand how the corrections modify the momentum structure of the dilepton. Thermal dilepton yields from the expanding QGP medium were determined by convoluting the rates with the space-time history of the collisions, along with the solutions of second-order causal hydrodynamics. We found that the modifications to the spectra due to total CE corrections are large at the early times and as the system evolves, they become minimal and converge to the first-order correction. We observed that the CE corrections are well behaved compared to the Grad's ansatz in thermal particle production.  
\par
Our results demonstrate that the effect of viscosity on experimentally relevant observables cannot be inferred solely from the magnitude of the underlying non-equilibrium correction. Instead, the observable dependence is controlled by the interplay between the momentum structure of the viscous correction and the kernel associated with the observable under consideration. This provides a unified framework for understanding the propagation of non-equilibrium effects from microscopic distribution functions to macroscopic observables in relativistic heavy-ion collisions.

\section*{Acknowledgements}
\par
L. J. N. acknowledges the Department of Science and Technology, Government of India for the
INSPIRE Fellowship.




\bibliography{reference}

\end{document}